\documentclass[12pt]{article}
\usepackage{amsmath}
\usepackage{graphicx}
\usepackage{enumerate}
\usepackage{natbib}

\usepackage{amssymb} 
\usepackage{mathrsfs} 
\usepackage{moreverb}
\usepackage{bm}
\usepackage{bbm}
\usepackage{mathtools}
\usepackage{xargs} 
\usepackage{hyperref}
\usepackage{xcolor} 
\usepackage{graphicx} 
\graphicspath{{figures/}} 
\usepackage{makecell} 
\usepackage{multicol} 
\usepackage{booktabs} 
\usepackage{subcaption} 
\usepackage[labelformat=parens,labelsep=quad,skip=3pt]{caption} 
\usepackage{float} 
\usepackage{ragged2e} 

\usepackage{xr} 
\newcommand{\blind}{1}

\begin{document}

	\def\spacingset#1{\renewcommand{\baselinestretch}%
		{#1}\small\normalsize} \spacingset{1}

	
	\if1\blind
	{
		\title{\bf Conditional Cross-Design Synthesis Estimators for Generalizability in Medicaid}
		\author{Irina Degtiar\textsuperscript{a}\thanks{Contact: Dr. Irina Degtiar, Department of Biostatistics, Harvard T.H. Chan School of Public Health, Boston, MA. (email: \href{mailto:idegtiar@g.harvard.edu}{idegtiar@g.harvard.edu}). This work was supported by NIH New Innovator Award  DP2MD012722
				and NIH  training grants T32LM012411 and T32ES07142.}, Tim Layton\textsuperscript{b}, Jacob Wallace\textsuperscript{c}, and Sherri Rose\textsuperscript{d} \hspace{.2cm}\\
			\small
			\textsuperscript{a}Harvard T.H. Chan School of Public Health\\
			\small
			\textsuperscript{b} Harvard Medical School\\
			\small
			\textsuperscript{c} Yale School of Public Health \\ 
			\small
			\textsuperscript{d} Stanford University}
		\maketitle
	} \fi

	\if0\blind
	{
		\bigskip
		\bigskip
		\bigskip
		\begin{center}
			{\LARGE\bf Conditional Cross-Design Synthesis Estimators for Generalizability in Medicaid}
		\end{center}
		\medskip
	} \fi
	
	\bigskip
	\begin{abstract}
		While much of the causal inference literature has focused on addressing internal validity biases, both internal and external validity are necessary for unbiased estimates in a target population of interest. However, few generalizability approaches exist for estimating causal quantities in a target population when the target population is not well-represented by a randomized study but is reflected when additionally incorporating observational data. To generalize to a target population represented by a union of these data, we propose a class of novel conditional cross-design synthesis estimators that combine randomized and observational data, while addressing their respective biases. The estimators include outcome regression, propensity weighting, and double robust approaches. All use the covariate overlap between the randomized and observational data to remove potential unmeasured confounding bias. We apply these methods to estimate the causal effect of managed care plans on health care spending among Medicaid beneficiaries in New York City. 
	\end{abstract}
	
	\noindent%
	\vfill

	\newpage

	\begin{centering}
		\section{BACKGROUND}\label{sec:background}
	\end{centering}
	
	When estimating causal effects, randomized data estimates often have unbiased causal effects for the population {represented by the study} (i.e., internal validity). However, these estimates may not reflect causal effects in the target population (i.e., external validity), and, furthermore, not represent subsets of the target population. Observational data may be more representative of the target population and, hence, have external validity, but they are potentially affected by unmeasured confounding. These challenges arise in settings ranging from clinical trials that exclude certain patient subsets \citep{prentice2005,lu2019} to policy evaluation studies that aim to inform deployment in a different population \citep{attanasio2003, kern2016}.  While much of the causal inference literature has focused on addressing internal validity biases, both internal and external validity are necessary for unbiased estimates.  
	
	Although  generalizability and transportability methods exist for extending inference from a randomized study to a target population, few leverage a combination of randomized and observational data to address each data source's shortcomings \citep{degtiar2021}. 
	Approaches that do combine individual-level randomized and observational data  face limitations when the target population doesn't fully overlap  with the randomized data and the observational data  have unmeasured confounding. Existing techniques extrapolate from the randomized data beyond their support \citep{attanasio2003,hill2011,kern2016}, assume the included observational data have no unmeasured confounding \citep{kern2016,lu2019}, or allow for unmeasured confounding but assume treatment effects are identical within strata of effect modifiers, which may not hold with continuous effect modifiers \citep{rosenman2020}.  
	Cross-design synthesis methods combine randomized and observation data, often relying on a binary flag that determines eligibility to be in the randomized subset of the data, which requires overlap membership to be known \citep{begg1992, kaizar2011, greenhouse2017}. 
	Bayesian calibrated risk-adjusted modeling, currently only deployed in the context of Cox proportional hazards survival regression, necessitates a third external data source that has strong overlap with both the randomized and observational data \citep{varadhan2016,henderson2017}. A 2-step regression approach by \cite{kallus2018} assumes that the randomized covariate distribution is subsumed in the observational data distribution and does not directly extend to estimating population treatment-specific means rather than average treatment effects.

	We present a novel class of  methods, which we refer to as conditional cross-design synthesis (CCDS) estimators,  that address several limitations of existing estimators that incorporate outcome information from  randomized and observational data. All CCDS approaches estimate a conditional bias term from the overlapping support between randomized and observational data that is then used to `debias' observational data estimates. These techniques are robust  to unmeasured confounding in the observational data and positivity violations for selection into the randomized data. The estimators include outcome regression, 2-step outcome regression, inverse probability weighting, and double robust augmented inverse probability weighting approaches.  Our implementation allows for the incorporation of ensemble machine learning to estimate the regression components of the various estimators, minimizing reliance on misspecified parametric regressions. 

	We apply our class of CCDS estimators to a study in New York City (NYC) Medicaid Managed Care (MMC), which provides health insurance to most New York Medicaid beneficiaries. Beneficiaries who do not choose a health plan are randomly assigned to one. However, the 7\% of NYC beneficiaries who are randomized are not representative of the broader NYC Medicaid population. Of the remaining 93\% of enrollees who actively chose their health plan (i.e., the observational beneficiaries), some are not well-represented by any randomized beneficiaries. This motivates our CCDS approaches that combine randomized and observational data to estimate health plan-specific causal effects on health care spending in the full NYC Medicaid population.
	
	Section \ref{sec:notation_estimand} defines notation and the estimand of interest. Section \ref{sec:assumptions_id} reviews standard generalizability assumptions, describes our relaxation of two of the assumptions through the combination of randomized and observational data, and identifies the estimand of interest under our relaxed assumptions. Section \ref{sec:estimators} presents the novel CCDS estimators and the limited available alternative approaches. We evaluate all estimators through a simulation study in Section \ref{sec:sim} that highlights settings where each CCDS estimator can be anticipated to perform well. Section \ref{sec:rwd} applies these methods to our NYC Medicaid study examining the impact of managed care plans on health care spending.  Section \ref{sec:discussion} concludes with a discussion.
	
	\begin{centering}
		\section{NOTATION AND ESTIMAND}\label{sec:notation_estimand}  
	\end{centering} 
	\subsection{Notation}
	The target population of interest is represented by a target sample. A portion of the target sample is randomized to the intervention (i.e., managed care plans) and the remaining individuals are observational. Hence the target sample is a union of randomized and observational data. We observe $n=n_\text{RCT}+n_\text{obs}$ independent draws from an underlying probability distribution $P\in \mathcal{M}$, where $\mathcal{M}$ is statistical model, namely, a collection of possible probability distributions. Each of these draws consist of an outcome $Y \in \mathbb{R}$, the intervention $A \in \mathcal{A}$, the vector of covariates $\bm{X} \in \mathcal{R} \in \mathbb{R}^d$, where $\mathcal{R}$ is the region of  support in the target population's covariate distribution, and an indicator for selection into the randomized group $S \in \mathcal{S} = \{0,1\}$. Thus, the observational unit for the target sample is $O = (Y,A,\bm{X},S)$.

	The data generating processes which result in randomized and observational data realizations differ. The randomized data consist of $n_\text{RCT}$ i.i.d.\ realizations conditional on selection into the randomized group, $S=1$. The observational unit for the randomized data is
	$O_\text{RCT} = (Y,A,\bm{X},S=1) \sim (Y,A,\bm{X}|S=1) \equiv P_\text{RCT}$.
	Similarly, the observational data consist of $n_\text{obs}$ i.i.d.\ draws conditional  on selection into the observational study, $S=0$. The observational unit for the observational data is thus $O_\text{obs} = (Y,A,\bm{X},S=0) \sim (Y,A,\bm{X}|S=0) \equiv P_\text{obs}$.
	
	\subsection{Estimand}
	As per the potential outcomes framework, let $Y^{a}$ be the potential outcome if intervention $a$ were assigned. The estimands of interest for our intervention are the target population treatment-specific means (PTSMs): $E(Y^{a}) \text{ for } \forall a \in \mathcal{A}$, as have been explored in prior analyses with multiple unordered treatments \citep{rose2019double}. In contrast, study treatment-specific means (STSMs) are mean counterfactual outcomes for a given treatment over a given study population:  $E(Y^{a}|S=s) \text{ for } \forall a \in \mathcal{A}, s \in \mathcal{S} $.  Because no given health plan serves as a natural ``control'' comparator, treatment-specific means rather than the target population average treatment effect (PATE: $E(Y^{a})-E(Y^{a'})$) are of interest.

	\subsection{Defining and Determining Overlap and Nonoverlap Regions}\label{sec:overlap}
	
	\begin{figure} 
		\spacingset{1}
		\centering
		\includegraphics*[width=0.5\textwidth, keepaspectratio=true]{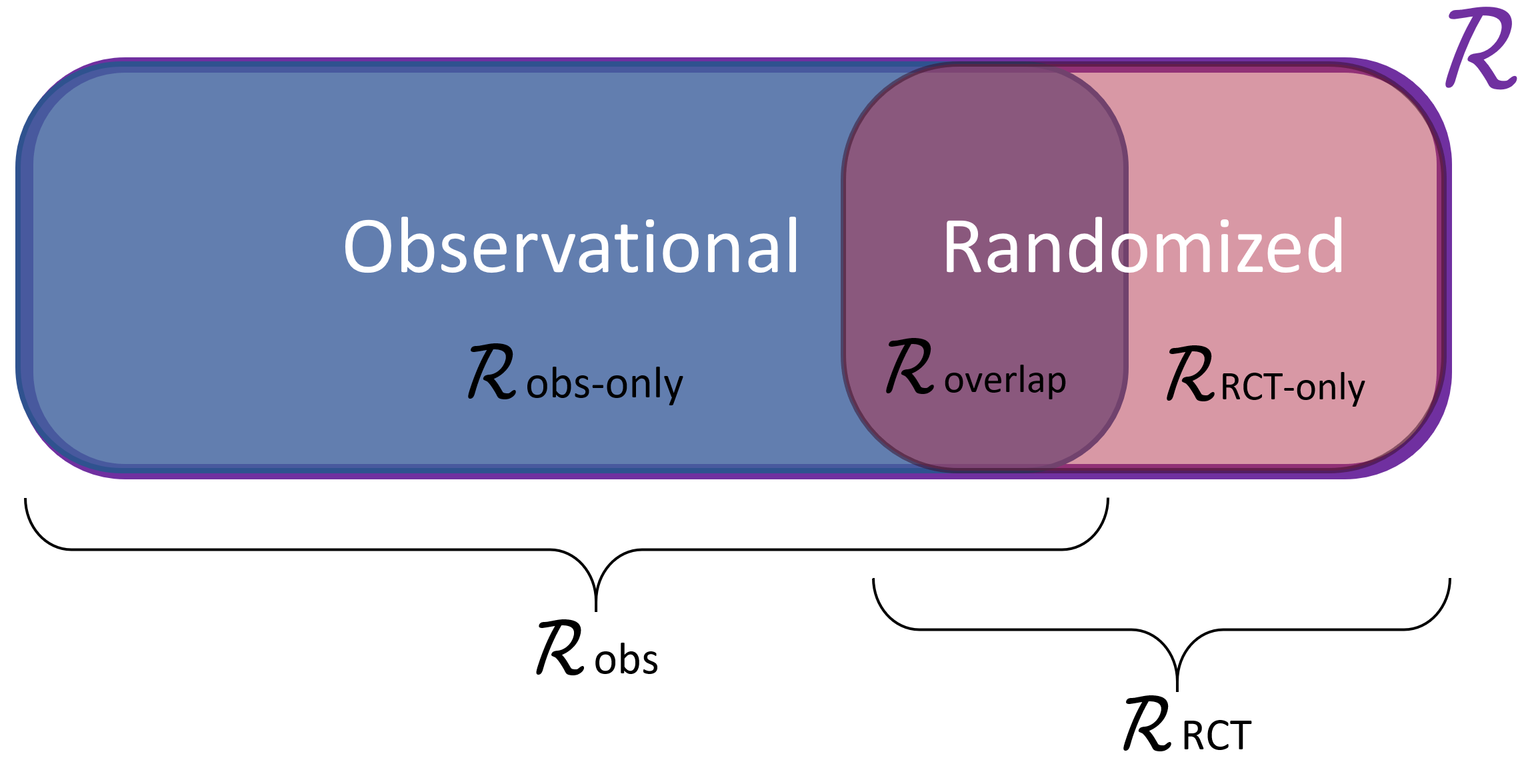}
		\caption[]{Overlap and Nonoverlap Regions in the Target Population. The region of support in the target population's covariate distribution, $\mathcal{R}$, is the union of the randomized group's support ($\mathcal{R}_\text{RCT}$) and the observational group's support ($\mathcal{R}_\text{obs}$). Partial overlap exists between $\mathcal{R}_\text{RCT}$ and $\mathcal{R}_\text{obs}$: $\mathcal{R}_\text{overlap}$ corresponds to the region of overlap (i.e., region of common support in the covariate distributions) between data sources; $\mathcal{R}_\text{obs-only}$ corresponds to the region only represented in the observational data and $\mathcal{R}_\text{RCT-only}$ corresponds to the region only represented in the randomized data.}
		\label{fig1}		
	\end{figure}

	Covariate distributions differ between randomized and observational groups: $P(\bm{X}|S=1) \ne P(\bm{X}|S=0)$. Furthermore $P(\bm{X}=\mathbf{x}|S=0)=0$ and $P(\bm{X}=\mathbf{x}'|S=1)=0$ for some $\mathbf{x},\mathbf{x}' \in \mathcal{R}$. Namely, a portion of the observational data is not well-represented in the randomized data and potentially a portion of the randomized data may not be well-represented in the observational data. However, there is a region of overlap between randomized and observational covariate distributions. Overlap refers to common support across randomized and observational populations in the distribution of outcome predictors associated with study selection (or effect modifiers associated with study selection if the estimand of interest had been an average treatment effect): $\mathcal{R}_\text{overlap} = {\mathbf{x} \in \mathcal{R} : P(\bm{X}=\mathbf{x}|S=1) > 0 \cap P(\bm{X}=\mathbf{x}|S=0) > 0}$ (Figure \ref{fig1}). Regions of nonoverlap therefore correspond to regions of the covariate distribution where either only observational individuals ($\mathcal{R}_\text{obs-only}$) or only randomized individuals ($\mathcal{R}_\text{RCT-only}$) would be observed, i.e., regions where units in one study population are not eligible to be in the other study population.

	The target sample covariate distribution ($\mathcal{R}$) can therefore be decomposed as: $\mathcal{R} = \mathcal{R}_\text{overlap} \cup \mathcal{R}_\text{obs-only} \cup \mathcal{R}_\text{RCT-only}$. Thus, $\mathcal{R}_\text{obs} = \mathcal{R}_\text{overlap} \cup \mathcal{R}_\text{obs-only}$ and $\mathcal{R}_\text{RCT} = \mathcal{R}_\text{overlap} \cup \mathcal{R}_\text{RCT-only}$. $\mathcal{R}_\text{RCT-only}$ and $\mathcal{R}_\text{obs-only}$ may be null sets. Let $R$ be an indicator for being in the respective region, e.g., $R_\text{overlap} = \mathbbm{1}$(membership in $\mathcal{R}_\text{overlap}$).

	At times, it may be the case that rather than a union of a randomized and observational study being representative of the target population, a reweighted union of the two studies may be representative, such as when working with a random sample of observational data for computational efficiency (which we do for our analysis), or when data are collected through survey sampling. In this case, through reweighting, one can map the randomized and observational study regions of covariate support, $\mathcal{R}_\text{RCT}$ and $\mathcal{R}_\text{obs}$, into a transformation, $\mathcal{R}_\text{RCT} \rightarrow \mathcal{R}_\text{RCT}^*$ and $\mathcal{R}_\text{obs} \rightarrow \mathcal{R}_\text{obs}^*$, in which the decomposition above of $\mathcal{R}^* = \mathcal{R}^*_\text{overlap} \cup \mathcal{R}^*_\text{obs-only} \cup \mathcal{R}^*_\text{RCT-only}$ holds. Note that this includes the possibility of the target population being represented by just the observational data.

	While the above definition of overlap corresponds to a population feature, nonoverlap can also occur due to having a finite sample; by chance, the data may be sparse in some region of the covariate distribution even though that region has support. In practice, we will account for overlap  as both a population and sample feature, determining regions of the covariate space that have common support and observed data from both groups.
	To estimate the region of overlap, $\mathcal{R}_\text{overlap}$, we extend a data-driven approach for determining areas of treatment overlap based on propensity scores for treatment assignment  \citep{nethery2018}.
	We adopt a similar approach for the propensity score for study selection $\pi_S = P(S|\bm{X})$, but on the logit scale to give more granularity to very low and very high propensity scores. The region of overlap consists of points in the logit of the propensity score for selection that have at least $\beta$ observations from each study group within an interval of size $\alpha$ around that point  \citep{nethery2018}.

	\begin{centering}
		\section{ASSUMPTIONS AND IDENTIFICATION}\label{sec:assumptions_id}  
	\end{centering}
	\subsection{Standard Assumptions}\label{sec:assumptions}
	Identifying population causal quantities such as PTSMs and PATEs standardly relies on the following sufficient generalizability assumptions \citep{stuart2011,tipton2013,degtiar2021}:
	
	\noindent \textbf{Internal validity} 
	\begin{enumerate}
		\item Conditional treatment exchangeability: $Y^{a}\bot A\ |\bm{X},S=s$ for all $a\in \mathcal{A}, s\in \mathcal{S}$.
		\item Positivity of treatment assignment: $P\left(\bm{X}=\bm{x}|S=1\right)>0\ \Rightarrow $ $P\left(A=a|\bm{X}=\bm{x},S=1\right)>0$ with probability 1 for all $a\in \mathcal{A}$.
		\item Stable unit treatment value assumption (SUTVA) for treatment assignment: if $A_i=a$ then $Y_i=Y^{a}_i$.
	\end{enumerate}
	\textbf{External validity}
	\begin{enumerate}
		\setcounter{enumi}{3}
		\item Conditional exchangeability for study selection: $Y^{a}\bot S\ |\bm{X}$ for all $a\in $ $\mathcal{A}$.
		\item Positivity of study selection: $P\left(\bm{X}=\bm{x}\right)>0\ \Rightarrow P\left(S=s|\bm{X}=\bm{x}\right)>0$ with probability 1 for all $s \in \mathcal{S}$.
		\item SUTVA for study selection: if $S_i=s$ and $A_i=a$ then $Y_i=Y^a_i$.
	\end{enumerate}
	
	\noindent For a specific estimand of interest, Assumptions 1 and 4 can be weakened. For example, when estimating PTSMs, Assumptions 1 and 4 can be replaced by the following:
	
	\begin{itemize}
		\item[1.] Mean conditional treatment exchangeability: $E(Y^{a}|A=a,S=s,\bm{X}) = E(Y^{a}|S=s,\bm{X})$ for all $a\in \mathcal{A}, s\in \mathcal{S}$.
		\item[4.] Mean conditional exchangeability for study selection: $E(Y^{a}|S=s,\bm{X}) = E(Y^{a}|\bm{X})$ for all $a\in $ $\mathcal{A}, s\in \mathcal{S}$.
	\end{itemize}

	\subsection{Relaxation of the Mean Conditional Treatment Exchangeability and Positivity of Study Selection Assumptions}
	To accommodate the potential violations of standard assumptions 1 and 5 for PTSMs, 
	we replace these assumptions with the following relaxations:
	
	\begin{itemize}
		\item[1b.] Mean conditional exchangeability in the randomized group: $$E(Y^{a}|S=1,A=a,\bm{X})=E(Y^{a}|S=1,\bm{X}) \text{ for all }a\in \mathcal{A},$$ 
		
		and constant conditional bias in the observational group:
		\begin{align*}
			E(Y^{a}|S=0,A=a,\bm{X})&-E(Y^{a}|S=1,A=a,\bm{X})  = \\
			E(Y^{a}|S=0,A=a,R_\text{overlap}=1,\bm{X})&-E(Y^{a}|S=1,A=a,R_\text{overlap}=1,\bm{X}).
		\end{align*}  
		
		\item[5b.] Overlap between study samples: there exists a non-null set $\mathcal{R}_\text{overlap}$ such that $P(\bm{X}=\bm{x}|R_\text{overlap})>0\ \Rightarrow P\left(S=s|\bm{X}=\bm{x}\right)>0$ with probability 1 for all $s\in \mathcal{S}$. 
	\end{itemize}
	
	Assumption 1b corresponds to the same conditional bias relationship holding in $\mathcal{R}_\text{overlap}$ as $\mathcal{R}_\text{obs}$: $b(a,\bm{x})=b(a,\bm{x}|R_\text{ overlap}=1)$, for all $a \in \mathcal{A}$ where $b(a,\bm{x}) \equiv E(Y^a|S=0,A=a,\bm{X}=\bm{x})-E(Y^a|S=0,\bm{X}=\bm{x})$. See Appendix \ref{appendix:assumptionDerivation} in the supplementary material for a derivation and further motivation for these weakened identifiability assumptions, in addition to a restatement of Assumption 1b with respect to the unmeasured confounders that are implicitly being integrated over.
	
	More specifically (and more weakly), Assumption 1b must hold in expectation over the $\bm{X}$ covariate distribution in the observational data (mean constant conditional bias):
	\begin{align*}
		E_{\bm{X}}\big\{E(Y^{a}|S=0,A=a,\bm{X})&-E(Y^{a}|S=1,A=a,\bm{X})\big|S=0\big\}  = \\
		E_{\bm{X}}\big\{E(Y^{a}|S=0,A=a,R_\text{overlap}=1,\bm{X})&-E(Y^{a}|S=1,A=a,R_\text{overlap}=1,\bm{X})\big|S=0\big\}.
	\end{align*}

	Assumption 1b states that the relationship between bias and measured covariates is unrelated to being in the overlap vs. nonoverlap regions, i.e., that the distribution of unmeasured confounders does not differ between $\mathcal{R}_\text{overlap}$ and $\mathcal{R}_\text{obs}$, conditioning on $\bm{X}$ and $A$. This assumption is strictly weaker than the no unmeasured confounding assumption in that the assumption of no unmeasured confounding is nested within Assumption 1b: with no unmeasured confounding, $b(a,\bm{x})=0$. Constant conditional bias can be seen as an extension of the assumption made for cross-design synthesis \citep{kaizar2011}, except that constant conditional bias is allowed to depend on measured covariates and a dichotimization of the covariate distribution support into overlap and nonoverlap regions replaces predefined eligibility determining overlap region membership. We hence assume that the covariates $\bm{X}$ capture all factors that would lead to differential bias in the overlap as nonoverlap regions. This suggests that we can estimate bias in the overlap region and use those estimates to extrapolate to and correct for bias in the observational group's nonoverlap region. 
	
	Assumption 1b is untestable, just as is the assumption of no unmeasured confounding; it would fail if the processes that drove unmeasured confounding differed between overlap and nonoverlap regions in a way that was not captured by measured covariates, or if the distribution of the unmeasured confounder differed between those regions in such a way as to create different conditional expectation relationships. This could occur, for example, if an unmeasured confounder drove overlap region membership. If the constant bias assumption is not reasonable for a given setting, one can alternatively perform sensitivity analysis to obtain bounds on PTSMs (Appendix \ref{appendix:sensAnalysis}).
	
	In practice, Assumption 5b's region of overlap should be sufficiently large to learn the bias term, i.e., sufficiently large for Assumption 1b to hold. Empirical violations of Assumption 5b are partially testable using $\pi_S$; the existence of overlap in the propensity score distributions between randomized and observational groups provides evidence for this assumption. 
	Observational group propensity scores may also be close to zero and lack overlap with randomized group propensity scores when the observational group size far exceeds the randomized group.

	Of note, the $\bm{X}$ needed for Assumptions 1b and 2 and the $\bm{X}$ needed for Assumption 4 and 5b 
	may differ. As a result, the region of overlap should exist with respect to outcome predictors but should be large enough to ensure that Assumption 1b holds. It is therefore reasonable to use an $\bm{X}$ matrix that contains all outcome predictors and confounders to assess all assumptions. Thus, as described earlier, our $\bm{X}$ is the union of the covariates sets needed for all assumptions to hold.

	\subsection{Identification}
	Under the modified assumptions above, the causal estimand of interest can be identified by the following CCDS functional of the observed data: 
	\begin{align*}
		\psi_\text{CCDS}(a)=&E_{\bm{X}|S=1}\Big[E(Y|S=1,A=a,\bm{X})\Big|S=1\Big] P(S=1) \\ &+ E_{\bm{X}|S=0}\Big[E(Y|S=0,A=a,\bm{X})\Big|S=0\Big]P(S=0) \\
		&- E_{\bm{X}|S=0}\Big[\big\{E(Y|S=0,A=a,R_\text{overlap}=1,\bm{X}) \\ & \quad  \quad 
			- E(Y|S=1,A=a,R_\text{overlap}=1,\bm{X})\big\}\Big|S=0\Big]P(S=0).   
	\end{align*}
	See Appendix \ref{appendix:identificationProof} for the proof and Appendix \ref{appendix:altDecomp} for alternative functionals that identify the PTSM, derived through different decompositions of the data.
	
	\begin{centering}
		\section{ESTIMATORS}\label{sec:estimators}  
	\end{centering}
	We develop four novel estimators that combine randomized and observational data to estimate PTSMs relying on our CCDS framework. The novel estimators consist of outcome regression, 2-stage outcome regression, inverse probability weighting, and double robust augmented inverse probability weighting approaches. 	
	
	\subsection{CCDS Outcome Regression Estimator}
	The CCDS outcome regression (CCDS-OR) estimator uses outcome regressions to estimate the combination of the conditional distributions in $\psi_\text{CCDS}(a)$:
	\begin{align*}
		\hat{\psi}_\text{CCDS-OR}(a) = &\frac{1}{n}\sum_{i=1}^{n} \hat{Q}(S_i=1,A_i=a,\bm{X}_i) \mathbbm{1}(S_i=1) +  \hat{Q}(S_i=0,A_i=a,\bm{X}_i) \mathbbm{1}(S_i=0)  \nonumber\\ &\quad \quad - \Big\{\hat{Q}(S_i=0,A_i=a,\hat{R}_\text{overlap, i}=1,\bm{X}_i) \\ & \qquad \quad \quad - \hat{Q}(S_i=1,A_i=a,\hat{R}_\text{overlap, i}=1,\bm{X}_i) \Big\}\mathbbm{1}(S_i=0),
	\end{align*}
	where $\hat{R}_\text{overlap}$ is estimated as described in Section \ref{sec:overlap}, $\hat{Q}(S=1,A=a,\bm{X})$ is an estimator for ${E}(Y|S=1,A=a,\bm{X})$, $\hat{Q}(S=0,A=a,\bm{X})$ is an estimator for ${E}(Y|S=0,A=a,\bm{X})$, $\hat{Q}(S=0,A=a,\hat{R}_\text{overlap}=1,\bm{X})$ is an estimator of ${E}(Y|S=0,A=a,R_\text{overlap}=1,\bm{X})$, and $\hat{Q}(S=1,A=a,\hat{R}_\text{overlap}=1,\bm{X})$ is an estimator of ${E}(Y|S=1,A=a,R_\text{overlap}=1,\bm{X})$. The first term corresponds to treatment specific mean estimates for the randomized subset of the target sample, the second term provides preliminary estimates for the observational subset of the target sample, and the third term debiases the preliminary observational data estimates. Implementation considerations for regression choices and a conditional treatment-specific mean version of the estimator are presented in Appendix \ref{appendix:implementation}.

	\subsection{2-stage CCDS Outcome Regression Estimator}
	To avoid overfitting to overlap region trends, the 2-stage CCDS estimator replaces the debiasing term, the third term, with a 2-stage regression:
	
	\begin{align*}
		\hat{\psi}_\text{2-stage CCDS}(a) = &\frac{1}{n}\sum_{i=1}^{n} \hat{Q}(S_i=1,A_i=a,\bm{X}_i) \mathbbm{1}(S_i=1) +  \hat{Q}(S_i=0,A_i=a,\bm{X}_i) \mathbbm{1}(S_i=0) \nonumber\\  &\quad \quad -  \hat{b}(S_i=1,a,\bm{X}_i) \mathbbm{1}(S_i=0),
	\end{align*}
	where $\hat{b}(S_i=1,a,\bm{X}_i)$ is estimated via
	\begin{itemize}
		\item[(1)] $\hat{b}'(S_i=1,a,\bm{X}_i) = \hat{Q}(S_i=0,A_i=a,\hat{R}_\text{overlap, i}=1,\bm{X}_i)\mathbbm{1}(S_i=1, \hat{R}_\text{overlap, i}=1) - \hat{Q}(S_i=1,A_i=a,\hat{R}_\text{overlap, i}=1,\bm{X}_i)\mathbbm{1}(S_i=1, \hat{R}_\text{overlap, i}=1),$
		\item[(2)] $\hat{b}'(S_i=1,a,\bm{X}_i)= \frac{\hat{w}_\text{bias}(S_i,\bm{X}_i)}{\sum_{i=1}^n\hat{w}_\text{bias}(S_i,\bm{X}_i)} \hat{g}(\bm{X}_i)$ with 
		$$\hat{w}_\text{bias}(S_i,\bm{X}_i)=\frac{\mathbbm{1}(S_i=1,\hat{R}_\text{overlap, i}=1)\hat{P}(S_i=0|\bm{X}_i)}{\hat{P}(\hat{R}_\text{overlap, i}=1|S_i=1,\bm{X}_i)\hat{P}(S_i=1|\bm{X}_i)},$$ and $\hat{g}(\bm{X})$ an estimator of a regression function described below. $\hat{b}(S_i=1,a,\bm{X}_i)$ is estimated for the observational data from this fixed weighted regression.
	\end{itemize}

	Namely, Stage (1), estimates an intermediate bias term $\hat{b}'(S_i=1,a,\bm{X}_i)$ using randomized overlap data: bias estimates are the difference in predicted counterfactual outcomes using regressions fit to the overlap region of the observational vs.\ randomized data, creating predictions for the randomized overlap data. As there is no bias in expectation in the randomized overlap data, any estimated bias stems from the regression $\hat{Q}(S_i=0,A_i=a,\bm{X}_i)$. Stage (2) then fits a weighted regression with the estimates of $\hat{b}'(S_i=1,a,\bm{X}_i)$ from Stage (1) as the outcome. This second stage focuses on the relationship between the bias estimates in the overlap region and measured covariates. The debiasing term $\hat{b}(S_i=1,a,\bm{X}_i)$ is then estimated using the observational data and the fixed $\hat{g}(\bm{X})$ fit in Stage (2).

	The weight, $\hat{w}_\text{bias}$, standardizes the randomized data to the observational data so that the bias term is  estimated for the covariate distribution of interest. The weights follow from $P(S=0) = E\big[P(S=0|\bm{X})\big] = E\big[\mathbbm{1}(S=1,R_\text{overlap}=1)P(S=0|\bm{X})/\big(P(R_\text{overlap}=1|S=1,\bm{X})P(S=1|\bm{X})\big)\big]$. Reweighting will frequently not face issues when positivity of selection violations occur because $S=1$ data are used to estimate the bias term  and thus should not have many values close to zero for $\hat{P}(S_i=1|\bm{X}_i)$, which is in the denominator of the weight. Thus, while weighting is not required in such a 2-stage approach, the weights add robustness compared to an unweighted approach without common drawbacks of weighting, such as variance inflation due to unstable weights.

	Appendix \ref{appendix:2stageEstimator} presents a 2-stage approach that does not restrict itself to the overlap region (2-stage whole data), which suffers from the same reliance on extrapolating beyond randomized group support as does using only the randomized data, highlighting the importance of focusing on the overlap region to debias observational data.

	\subsection{CCDS Inverse Probability Weighting Estimator}
	The cross-design synthesis inverse probability weighting (CCDS-IPW) estimator with stabilized weights uses propensity models to estimate PTSMs (see Appendix \ref{appendix:proofCCDSIPW} for the proof):  
	
	\begin{align*}
		\hat{\psi}_\text{CCDS-IPW}(a) &= \frac{n_\text{rand}}{n} \Bigg[\sum_{i=1}^n \hat{w}_{1}(S_i,A_i,\bm{X}_i)\Bigg]^{-1} \sum_{i=1}^n \hat{w}_{1}(S_i,A_i,\bm{X}_i)Y_i 
		\\ &+ \frac{n_\text{obs}}{n}\Bigg[\sum_{i=1}^n \hat{w}_{2}(S_i,A_i,\bm{X}_i)\Bigg]^{-1} \sum_{i=1}^n \hat{w}_{2}(S_i,A_i,\bm{X}_i)Y_i \nonumber\\
		&\quad -  \frac{n_\text{obs}}{n} \Bigg\{\Bigg[\sum_{i=1}^n \hat{w}_{3}(S_i,A_i,\bm{X}_i)\Bigg]^{-1} \sum_{i=1}^n \hat{w}_{3}(S_i,A_i,\bm{X}_i)Y_i \\
		& \qquad \quad  \quad - \Bigg[\sum_{i=1}^n \hat{w}_{4}(S_i,A_i,\bm{X}_i)\Bigg]^{-1} \sum_{i=1}^n \hat{w}_{4}(S_i,A_i,\bm{X}_i)Y_i
		\Bigg\},
	\end{align*}
	where:
	\begin{align*}
		\hat{w}_{1}(S_i,A_i,\bm{X}_i) &= \frac{\mathbbm{1}(S_i=1,A_i=a)}{\hat{P}(A_i=a|S_i=1,\bm{X}_i)}, \\
		\hat{w}_{2}(S_i,A_i,\bm{X}_i) &= \frac{\mathbbm{1}(S_i=0,A_i=a)}{\hat{P}(A_i=a|S_i=0,\bm{X}_i)}, \\
		\hat{w}_{3}(S_i,A_i,\bm{X}_i) &= \frac{\mathbbm{1}(S_i=0,A_i=a,\hat{R}_\text{overlap, i}=1)}{\hat{P}(\hat{R}_\text{overlap, i}=1|S_i=0,\bm{X}_i)\hat{P}(A_i=a|S_i=0,\hat{R}_\text{overlap, i}=1,\bm{X}_i)}, \\
		\hat{w}_{4}(S_i,A_i,\bm{X}_i) &= \frac{\mathbbm{1}(S_i=1,A_i=a,\hat{R}_\text{overlap, i}=1)[1-\hat{P}(S_i=1|\bm{X}_i)]}{\hat{P}(S_i=1|\bm{X}_i)\hat{P}(\hat{R}_\text{overlap, i}=1|S_i=1,\bm{X}_i)\hat{P}(A_i=a|S_i=1,\hat{R}_\text{overlap, i}=1,\bm{X}_i)} .
	\end{align*}
	Here, positivity of selection violations will usually not lead to unstable weights since $\hat{P}(\hat{R}_\text{overlap, i}=1|S_i=1,\bm{X}_i)\hat{P}(S_i=1|\bm{X}_i)$ only appears in the denominator for $\hat{w}_4$; these individuals, by overlap region construction, have propensity scores for selection bounded away from zero. Normalizing weights by their sum adds stability \citep{robins2000}. Nonetheless, this method can face lack of efficiency and potentially unstable estimates, particularly from estimating the second bias term contribution weighted by $\hat{w}_4$, as the components are estimated using small subsets of the data relative to the overall sample---only individuals randomized in the overlap region on a given treatment arm. This problem is exacerbated with many treatment groups, particularly for rare treatments.

	\subsection{CCDS Augmented Inverse Probability Weighting Estimator} 
	Our double robust estimator provides consistent estimates  when either the outcome regressions or product of propensity regressions are consistently estimated in each of the terms of $\psi_\text{CCDS}(a)$. The CCDS augmented inverse probability weighted (CCDS-AIPW) estimator is as follows:
	\begin{align*}
		\hat{\psi}&_\text{CCDS-AIPW}(a) \nonumber\\&=\frac{1}{n}\sum_{i=1}^n \frac{n_\text{rand}}{n}\frac{\hat{w}_{1}(S_i,A_i,\bm{X}_i)}{\sum_{i=1}^n \hat{w}_{1}(S_i,A_i,\bm{X}_i)}\big\{Y_i-\hat{Q}_i(S=1,A=a,\bm{X})\big\}+\mathbbm{1}(S_i=1)\hat{Q}_i(S=1,A=a,\bm{X})  \\
		&\quad + \frac{n_\text{obs}}{n}\frac{\hat{w}_{2}(S_i,A_i,\bm{X}_i)}{\sum_{i=1}^n \hat{w}_{2}(S_i,A_i,\bm{X}_i)}\big\{Y_i-\hat{Q}_i(S=0,A=a,\bm{X})\big\}+\mathbbm{1}(S_i=0)\hat{Q}_i(S=0,A=a,\bm{X}) \\ 
		&\quad-\frac{n_\text{obs}}{n}\frac{\hat{w}_{3}(S_i,A_i,\bm{X}_i)}{\sum_{i=1}^n \hat{w}_{3}(S_i,A_i,\bm{X}_i)}\big\{Y_i-\hat{Q}_i(S=0,A=a,\hat{R}_\text{overlap}=1,\bm{X})\big\} \nonumber\\ & \qquad - \mathbbm{1}(S_i=0)\hat{Q}_i(S=0,A=a,\hat{R}_\text{overlap}=1,\bm{X}) \\
		&\quad+\frac{n_\text{obs}}{n}\frac{\hat{w}_{4}(S_i,A_i,\bm{X}_i)}{\sum_{i=1}^n \hat{w}_{4}(S_i,A_i,\bm{X}_i)}\big\{Y_i-\hat{Q}_i(S=1,A=a,\hat{R}_\text{overlap}=1,\bm{X})\big\} \nonumber\\ & \qquad  \quad +\mathbbm{1}(S_i=0)\hat{Q}_i(S=1,A=a,\hat{R}_\text{overlap}=1,\bm{X}),  
	\end{align*}
	with $\hat{w}(S_i,A_i,\bm{X}_i)$ and $\hat{Q}_i(S,A,\bm{X})$ as defined above. CCDS-AIPW is a double robust estimator that is asymptotically efficient when the propensity  and outcome regressions are estimated consistently. See Appendix \ref{appendix:IF} for a derivation of the efficient influence function.

	\subsection{Inference}
	Confidence intervals and standard errors in our machine-learning-based analyses were calculated using a nonparametric bootstrap \citep{efron1994}. When using parametric regressions, a sandwich variance approach can be used to derive sampling variance, following M-estimation theory. 
	
	\subsection{Comparison estimators}\label{sec:comparison_estimators}
	No existing methods address both the overlap and unmeasured confounding challenges specific to our data setting. While the estimator of \cite{rosenman2020} addresses overlap and unmeasured confounding, it assumes that treatment effects are identical between randomized and observational groups within the same stratum of effect modifiers, which is unlikely to hold in our setting. We therefore compare against two simple approaches.  The first (rand estimator) fits an outcome regression using randomized data to extrapolate to the entire target population, including outside its region of support \citep{kern2016}: $\hat{\psi}_\text{rand}(a) = 1/n\sum_{i=1}^{n} \hat{Q}(S_i=1,A_i=a,\bm{X}_i)$. This extrapolation may yield bias when the relationship between covariates and potential outcomes differs in $\mathcal{R}_\text{overlap}$ compared to $\mathcal{R}_\text{obs}$ in a way that cannot be extrapolated from the randomized data.  
	
	The second (obs/rand estimator) is similar  to \cite{kern2016} and \cite{prentice2006}, though those estimators fit one outcome regression to both randomized and observational data and estimate effects for either just the observational data or just the randomized data. The obs/rand estimator we deploy here fits an outcome regression using randomized data to estimate counterfactuals for the randomized data and fits an outcome regression using observational data to estimate counterfactuals for the observational data: $\hat{\psi}_\text{obs/rand}(a) = 1/n\sum_{i=1}^{n} \hat{Q}_i(S=1,A=a,\bm{X}) \mathbbm{1}(S_i=1) + \hat{Q}_i(S=0,A=a,\bm{X}) \mathbbm{1}(S_i=0)$. This approach assumes there is no unmeasured confounding in the observational data. We used outcome regressions for rand and obs/rand estimators rather than approaches that incorporate propensities for selection, as the latter will result in denominators close to zero due to lack of overlap. See Appendix \ref{appendix:lu2019} for AIPW versions of the rand and obs/rand estimators, presented in \cite{lu2019}.

	\begin{centering}
		\section{SIMULATION STUDIES}\label{sec:sim}  
	\end{centering}
	We designed a broad series of simulations to evaluate the finite sample performance of our novel CCDS estimators compared to alternative approaches for estimating PTSMs as well as the PATE, examining two treatment groups $A \in \{1,2\}$. We assessed performance of these estimators in the presence of (1) complex data-generating mechanisms such that the randomized data do not extrapolate well outside their support, (2) unmeasured confounding in the observational data, and (3) positivity of selection violations. We also studied alternative data generating processes including different sample sizes, constant bias violations, unmeasured confounding settings, overlap settings, ratios of $n_\text{RCT}$ to $n_\text{obs}$, positivity of selection violation settings, exchangeability of study selection violations, overlap region determination settings, propensity for selection relationships, alternative outcome models, and alternative regression fits. In total, we examined 84 different data generating scenario $\times$ regression choice combinations.
	
	In the base case, we generated a target population of 1 million individuals from which we drew random samples of size $n=10,000$, with data-generating mechanism $P(Y,S,A,\bm{X},U) =$ $P(\bm{X})P(U|\bm{X})P(S|\bm{X},U)P(A|S,\bm{X},U)P(Y|S,A,\bm{X},U)$. The data had four independent measured confounders $X_1,...,X_4 \sim N(0,1)$; an unmeasured confounder $U \sim Binom(0.5)$; selection into the randomized group driven by the strongest confounder such that there existed $\mathcal{R}_\text{RCT-only}$ ($S=1 \text{ if } X_1 > QNnorm(0.9)$), $\mathcal{R}_\text{obs-only}$ ($S=0 \text{ if }  X_1 < QNorm(0.5)$), and $\mathcal{R}_\text{overlap}$ ($S \sim Binom(0.5)$, otherwise). This study selection process resulted in approximately a 1:4 ratio of randomized to observational individuals. Treatment assignment was $A \sim Binom(0.6)$ for $S=1$ and $A \sim Binom(\text{logit}^{-1}(-0.8+0.125X_1+0.1X_2+0.075X_3+0.05X_4+0.1(X_1+1)^3+0.625U))$ for $S=0$. The outcome was generated from the same distribution for both groups, $Y \sim Norm(\mu_Y,1), \text{ where } \mu_Y=-1.5-3A+4X_1+4X_2+3X_3+2X_4+0.4(X_1+1)^3+4AX_1+10U$. 
	
Estimators were fit with linear outcome regressions as well as an ensemble of 8 machine learning approaches. We implemented 2000 simulation iterations and 1000 bootstrap replications to generate confidence intervals. Propensities and their products used in weight denominators were trimmed at 0.001. We implemented the simulations in R, including the \texttt{SuperLearner} package \citep{polley2019} and the \texttt{pw\_overlap} function for overlap region estimation \citep{nethery2018}. See Appendix \ref{appendix:simulation} for the correspondence of our simulation design with identifiability assumptions, descriptions of alternative data-generating mechanisms, and further implementation details.      Our code is available on GitHub: \href{https://github.com/idegtiar1/CCDS}{https://github.com/idegtiar1/CCDS}.

	\begin{figure}[t!]
		\spacingset{1}
		\begin{center}
			\includegraphics[width=0.8\textwidth]{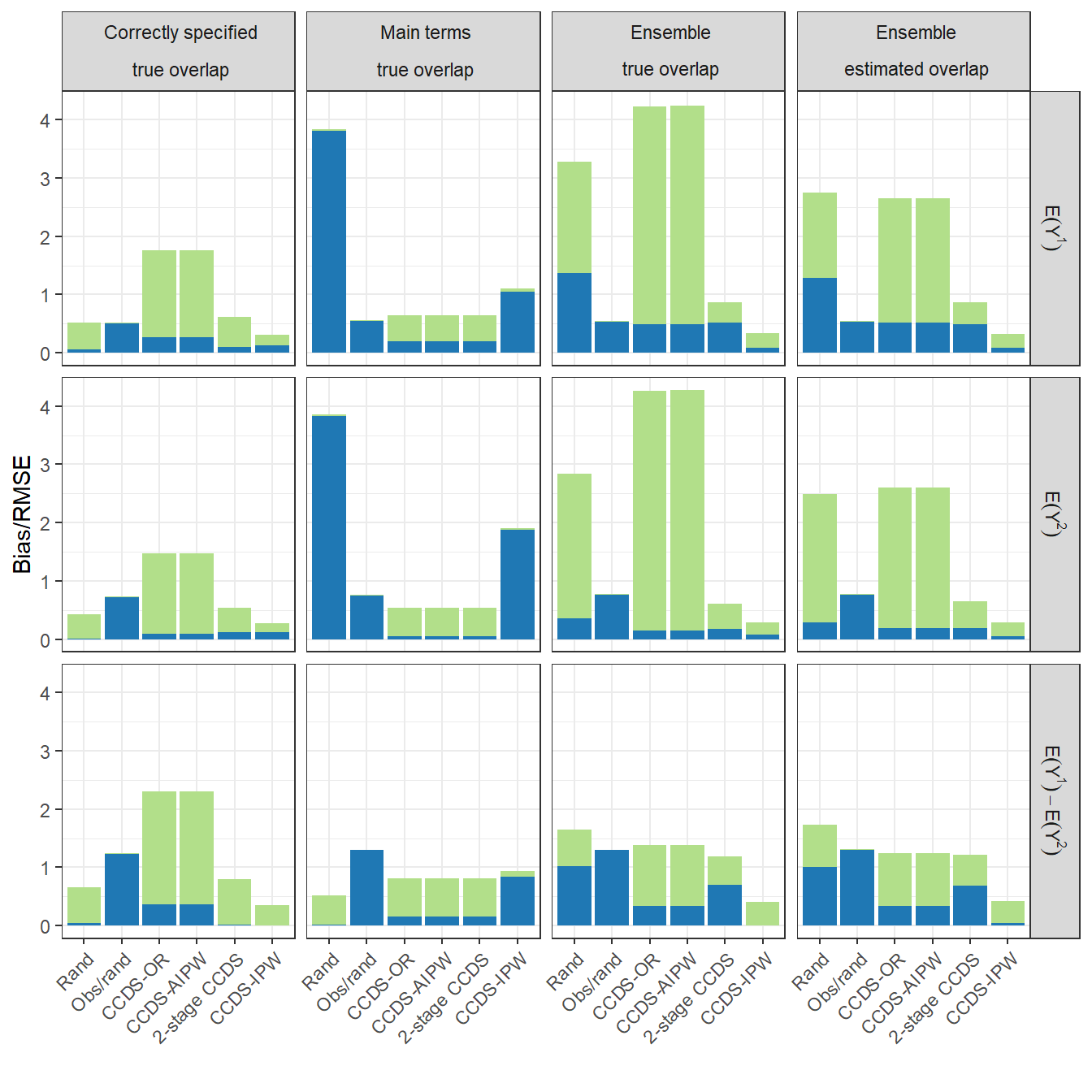}	
			\caption{Bias and RMSE for PTSM and PATE Estimates for $n=10,000$ Across 2000 Simulation Iterations and 1000 Bootstrap Replications. Absolute bias is the darker portion of each bar; RMSE corresponds to the total bar size.}		
			\label{figure:results_base_case}		
		\end{center}
	\end{figure}
	
	\textbf{Main Findings.} 	Results across different regression specifications highlight the estimators' relative strengths and disadvantages (Figure \ref{figure:results_base_case}). At the base case sample size of $n=10,000$, CCDS-OR and CCDS-AIPW performance was almost identical. These estimators suffered from large variability when fitting complex regressions in a small overlap region, which we observed in the correctly specified and ensemble settings. In contrast, the CCDS-OR and CCDS-AIPW estimators showed little bias and variance when fitting underspecified main terms regressions (underspecification avoids overfitting in a small overlap region). The 2-stage CCDS estimator decreased bias and variance when using correctly specified or ensemble regressions, relative to the (1-stage) CCDS-OR estimator and CCDS-AIPW. In the main terms setting, its estimates were identical to those of CCDS-OR due to linearity and additivity.

	The CCDS-IPW had the smallest bias and RMSE throughout all settings, except when fitting main terms regressions where it grossly misspecifies the propensity for selection, resulting in large remnant bias for that setting. However, the estimator's superior performance was due to the outcome model having more variability compared to the propensity models; e.g., the propensity for selection was deterministically assigned by $X_1$. With a more probabilistic relationship and smaller propensity scores, CCDS-IPW's bias increased (Appendix \ref{appendix:simulation_s_probabilistic}).

	While the rand estimator performed well with correctly specified regressions, using only main terms regressions resulted in large bias  due to poor extrapolation beyond the randomized data support. With more flexible ensemble approaches, the rand estimator suffered from both large bias and large variance. The obs/rand estimator was subject to unmeasured confounding bias, which was present even when correctly specified regressions were fit, though it had relatively low RMSE due to the large observational sample size. Results using \cite{lu2019}'s AIPW-based rand and obs/rand estimators had similar results to the outcome regression-based rand and obs/rand estimators, although the bias of \cite{lu2019}'s obs/rand estimators tended to be larger. This may be due to the misspecification of both outcome and propensity regressions and using less data to fit each regression due to sample splitting (Appendix Figure \ref{figure:full_sim_results}).

	\textbf{Estimating Overlap.} The last column of Figure \ref{figure:results_base_case} presents results from overlap region estimation using $\alpha=0.01 \times \text{range}(\text{logit}(\pi_S))$ and $\beta = 0.01 \times \text{min}(n_\text{RCT},n_\text{obs})$. With these specifications, compared to the truth, the estimated overlap region had a similar number of observational individuals (38\% vs. 35\%) and randomized individuals (50\% vs. 48\%). Performance was similar or better when estimating the overlap region in this setting and across the various other data-generating mechanisms and overlap region hyperparameter specifications we examined (Appendix \ref{appendix:simulation}).

	\begin{figure}[t!]
		\spacingset{1}
		\begin{center}		
			\includegraphics[width=1\textwidth]{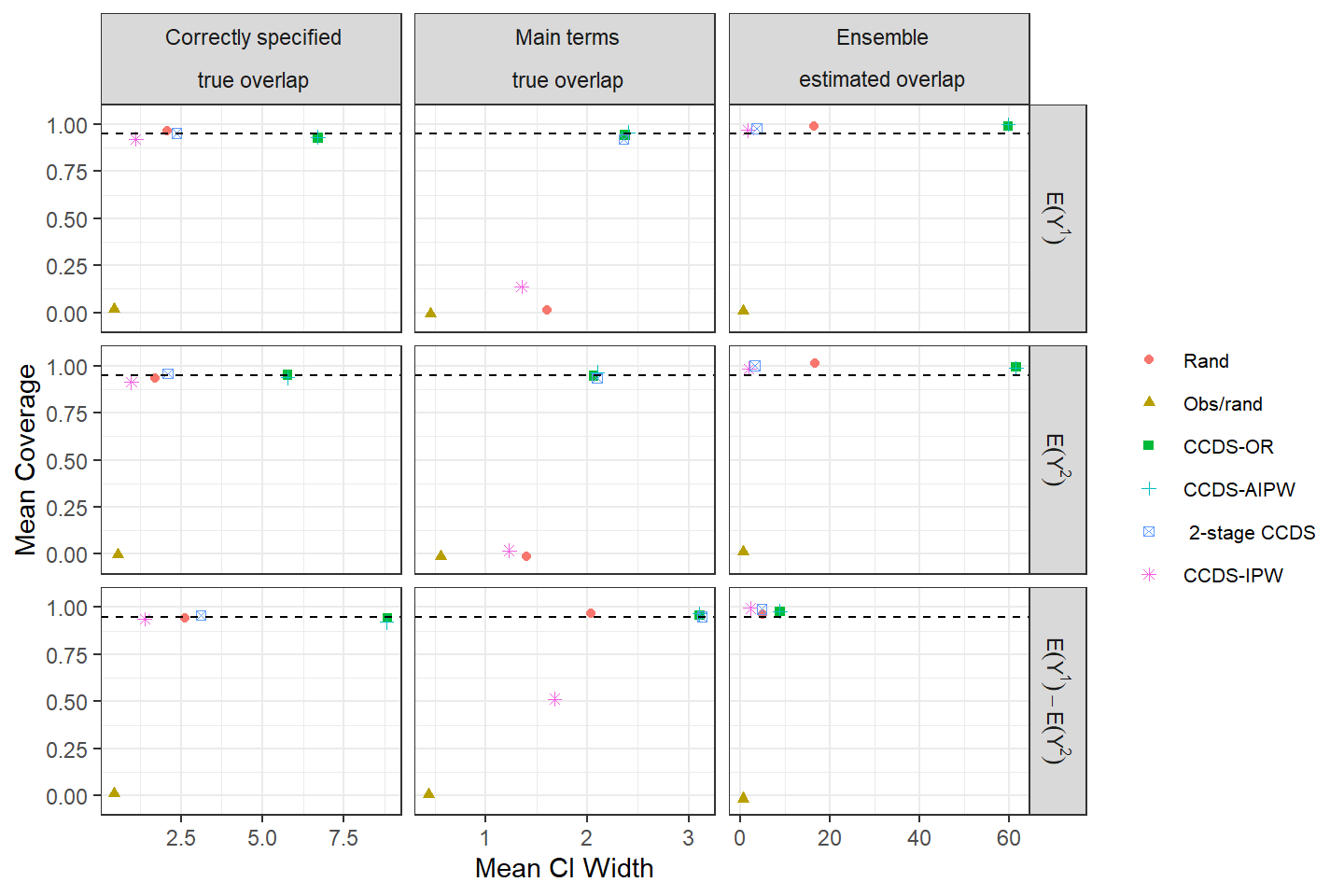}	
			\caption{Coverage and Confidence Interval Width for PTSM and PATE Estimates for $n=10,000$ Across 2000 Simulation Iterations and 1000 Bootstrap Replications. The dashed line corresponds to the target coverage of 95\%.}
			\label{figure:results_coverge}
		\end{center}
	\end{figure}
	
	\textbf{Coverage.} The obs/rand estimator showed 0\% coverage across all settings while all CCDS estimators were able to achieve nominal coverage, except the CCDS-IPW estimator when using grossly misspecified linear regressions (Figure \ref{figure:results_coverge}). The rand estimator attained 0\% coverage in the main terms settings for the PTSMs but 95\% coverage for the PATE, due to linear regressions correctly specifying the treatment effects but not the treatment-specific means in this data-generating mechanism; coverage remains low when the PATE does not extrapolate well  from the randomized data. Thus, while the bias and RMSE of the CCDS estimators may or may not  decrease compared to the obs/rand estimator (as shown in Figure~\ref{figure:results_base_case}) due to remnant estimation error from misspecifying regressions, which is particular evident with ensemble approaches, the poor coverage of the obs/rand estimator indicates this can be a false indication of precision.

	\textbf{Alternative Data-Generating Mechanisms.} CCDS estimator bias and RMSE shrunk with more overlap and with increasing proportions of randomized data. As unmeasured confounding bias increased, there was no corresponding increase in bias across CCDS estimators with correctly specified regressions and only a slight increase with ensembles. However, variance increased, reflecting additional uncertainty in settings with more unmeasured confounding. Violating the constant conditional bias assumption increased bias for the rand and all CCDS estimators, with CCDS estimators generally performing better than the rand estimator. All estimators performed poorly when the exchangeability of study selection assumption was violated. The RMSE for CCDS-IPW was most impacted by a smaller ratio of randomized to observational individuals. Overall, results for the CCDS estimators were similar across alternative data-generating mechanisms. Further details can be found in Appendix \ref{appendix:simulation}.

	\begin{centering}
		\section{MEDICAID STUDY}\label{sec:rwd}  
	\end{centering}

	Medicaid, administered by the Centers for Medicare \& Medicaid Services, provides insurance for low-income and disadvantaged Americans, covering a fifth of all individuals in the United States \citep{cms2020}. As described earlier, MMC provides health insurance plans for all but certain exempt groups \citep{medicaid2020}, and beneficiaries who do not actively choose a health plan are randomized to one. Understanding the impact of these individual MMC health plans on health care spending is an open question.  However, generalizing  the 7\% of beneficiaries who are randomized to the full NYC Medicaid population may be hampered by a lack of overlap in parts of the covariate distributions between randomized and observational (active chooser) groups. Yet, data from observational beneficiaries may be subject to potential unmeasured confounding from variables not captured in the claims data.

	We estimated the causal effects of enrollment into  NYC MMC health plans on health care spending for all NYC Medicaid beneficiaries with at least 6 months of follow-up, applying our novel CCDS and comparison estimators. 
Health care spending was examined  over 6 months on the log scale, as log(spending + 1), adjusting for baseline spending decile, age, documented sex, aid group, whether the beneficiary received social security income, neighborhood, and neighborhood poverty level. Further descriptions of the data can be found in \cite{geruso2020}. We used all 65,591 randomized beneficiaries and a 10\% random subset of observational beneficiaries within the study period (2008 - 2012) for computational efficiency, which totaled 98,232. Baseline spending was missing for 1\% of beneficiaries and was imputed to be zero (the most likely reason for missingness was no spending) along with an indicator for missingness. Regressions were fit using a \texttt{SuperLearner} ensemble (of \texttt{glm}, \texttt{glmnet} with $\alpha=0.5$, \texttt{gam}, and \texttt{nnet}). Propensity scores and their products used in weight denominators were trimmed at 0.001. To assess simultaneous 95\% coverage, a conservative Bonferroni adjustment was made to the bootstrap confidence intervals, which used 500 replications: each marginal confidence interval was constructed at the $1-0.05/k$ level, where $k=10$ plans. 
	
	Compared to randomized beneficiaries, observational beneficiaries differed across all measured factors: the latter were slightly younger (34.3 vs. 35.5 years old), spent less at baseline (\$2796 vs. \$3052), were more likely to have a documented sex of female (59\% vs. 40\%), were less likely to live in Manhattan (13\% vs 20\%) and more likely to live in Queens (28\% vs. 19\%), came from different aid groups, and were less likely to be eligible for social security income (2\% vs 9\%) (Appendix Table \ref{table:MedicaidTable1} in Appendix \ref{appendix:Medicaid}). Effect heterogeneity within the randomized data was driven by aid group status, supplemental security income eligibility, and neighborhood effects; within the observational data it was driven by neighborhood effects and receiving aid for children, all of which were imbalanced across randomized and observational beneficiaries, highlighting the need for generalizability approaches. 
	
	Overall, across all measured covariates, observational beneficiary characteristics were imbalanced across health plans, and these characteristics were also associated with health care spending, providing empirical evidence that these variables may be  confounders. While randomized beneficiaries were not representative of their observational counterparts, there was considerable covariate overlap, as measured by the propensity score for selection into the randomized subset of the data, though overlap was weakest where the observational data were most concentrated (Appendix Figure \ref{figure:results_Medicaid_piS} in Appendix \ref{appendix:Medicaid}). Using the conservative overlap hyperparameters $\alpha=0.01 \times \text{range}(\text{logit}(\pi_S))$ and $\beta=0.01 \times n_\text{RCT}$ resulted in 60\% of the target sample within the overlap region. The standardized mean difference in the propensity score for selection was 1.1 standard deviations, which far exceeds 0.25, one proposed threshold indicating large extrapolation \citep{stuart2011}, and, thus, supportive of the need for CCDS estimators.

	\begin{figure}[t!]
		\spacingset{1}
		\begin{center}
			\includegraphics[width=0.9\textwidth]{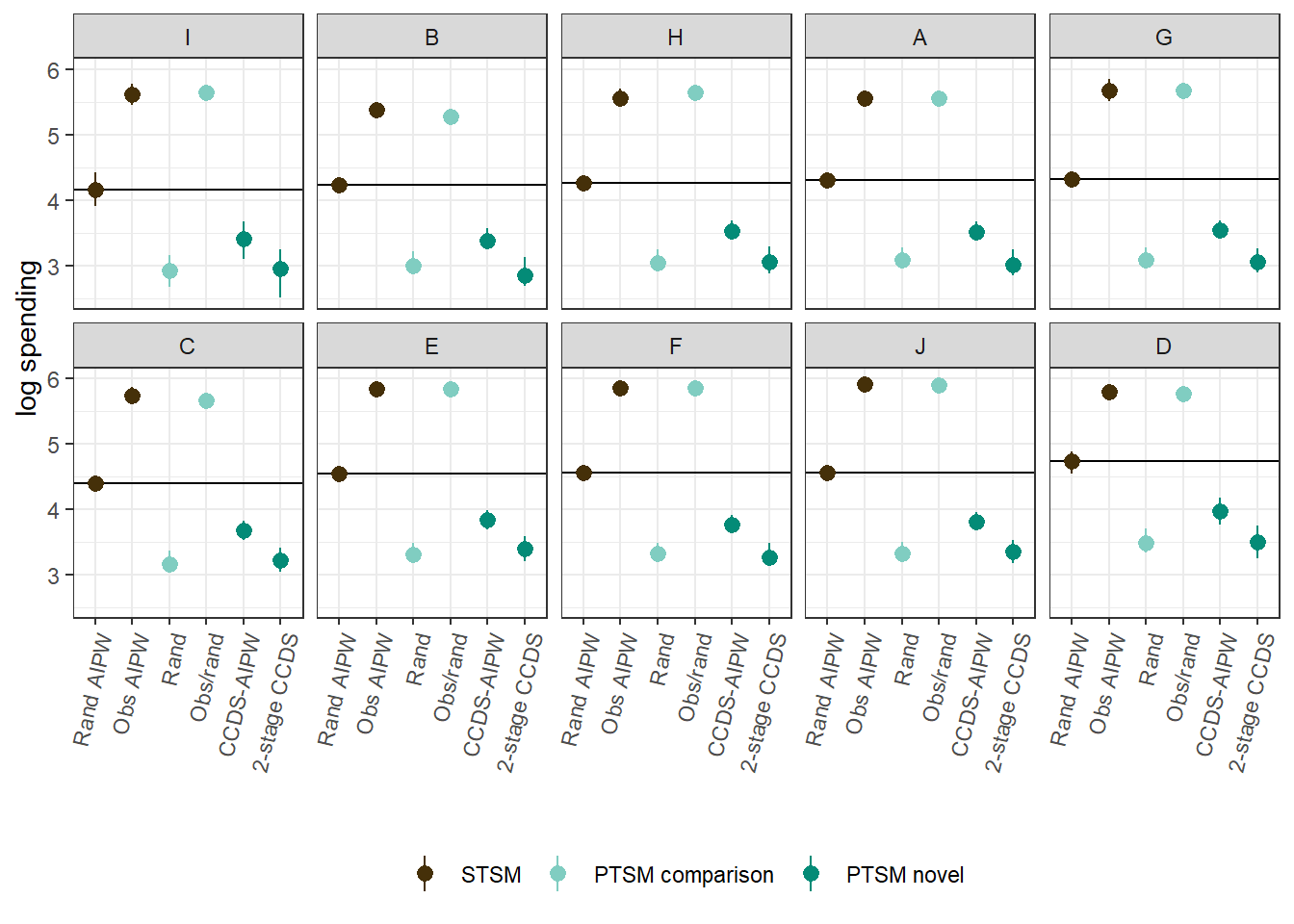}			
		\end{center}
		\caption{STSMs and PTSMs across Managed Care Plans, with 95\% Multiplicity-Adjusted Confidence Intervals. Plans are ordered by randomized data STSM estimates.}	
		\label{figure:results_Medicaid}	
	\end{figure}

	Figure \ref{figure:results_Medicaid} presents STSMs for the randomized and observational study populations and PTSMs for the NYC Medicaid target population, including results for  two CCDS estimators well suited to this setting. (All estimators are available in Appendix Figure \ref{figure:results_Medicaid_all} in Appendix \ref{appendix:Medicaid}.) Despite higher unadjusted mean spending in the randomized group, causal estimates of STSMs in the observational data were consistently higher than estimates of STSMs in the randomized data across all health plans. This discrepancy reflects both differences in population characteristics as well as potential unmeasured confounding in the observational data; neither estimate aligned with rand or CCDS estimates of PTSMs, which were consistently lower than randomized and observational STSMs. Results remained similar when accounting for country-month-year correlation (Appendix \ref{appendix:CMY}).
	
	Given the substantial overlap between randomized and observational covariate distributions, it is not surprising that CCDS estimates were in a similar range  to the rand estimates of PTSMs. However, the double robust CCDS-AIPW estimates were higher than rand estimates (12.5-16.7\% difference in log spending) and confidence intervals were non-overlapping for all but plans D and I. CCDS-AIPW did not show large variability with ensemble regressions, unlike in the simulations. Obs/rand estimates and observational data AIPW STSMs (which largely aligned as the observational data comprised 93\% of the data) were widely discrepant from other PTSMs, suggesting a large amount of unmeasured confounding bias in the observational data. Unlike in the simulation, CCDS-IPW confidence intervals were wider than those of other CCDS estimators, which is common to IPW estimators in practice, and also reflects the difficulty of estimating propensities for multiple treatments (Appendix Figure \ref{figure:results_Medicaid_all}). While the rand PTSM estimator could provide reasonable estimates in this setting, where there is a fair amount of overlap between randomized and observational data, the CCDS estimators were able to incorporate all data and did not rely on extrapolating spending estimates beyond the support of the randomized data.

	\begin{centering}
		\section{DISCUSSION}\label{sec:discussion}  
	\end{centering}
	When observational and randomized data are both available, there is potential to overcome each data type's limitations through their combination. Namely, when some individuals in the target population are not well-represented in the randomized data and the observational data have unmeasured confounding, neither data type alone can successfully generalize to the target population represented by a union of randomized and observational data. This article proposes a class of novel estimators that can surmount positivity of selection assumption violations in the randomized data and unmeasured confounding in the observational data by using common support between the data sources to remove unmeasured confounding bias. 
	
	The proposed outcome regression, propensity score, and double robust CCDS estimators have varying strengths.  When the functional forms of the true data generating processes can be approximated by simple linear regressions, the double robust CCDS-AIPW estimator with linear regressions is a suitable default approach for combining randomized and observational data. Even when linear regressions do not capture the full complexity of the data generating process, simulations showed that CCDS-AIPW and CCDS-OR with main terms regressions were able to recover unbiased estimates. However, when fitting more complex regressions, these estimators may lead to unstable bias extrapolations from the overlap region, although we did not see this drawback in our NYC Medicaid data analysis, which had a larger area of overlap. When more complex regression approaches are used, the 2-stage CCDS or CCDS-IPW may also be suitable, depending on whether there is more knowledge of the outcome relationship or the propensity for selection and treatment relationships and whether selection or treatments are rare or multinomial with small probabilities. The 2-stage approach improves performance compared to the CCDS-OR estimator by stabilizing initial estimates to alleviate overfitting to overlap region trends.

	In the NYC Medicaid data, the study and target population causal estimates were markedly different. Novel and existing generalizability methods helped reconcile these discrepancies by specifying a target population for which inference was desired. There were also significant differences between PTSM rand and obs/rand estimates, showcasing the need to account for both potentially poor extrapolation from the randomized data and potential unmeasured confounding in the observational data. The proposed CCDS estimators provided evidence that the observational data remained subject to unmeasured confounding bias even after adjusting for measured factors.
	
	Our CCDS framework is sensitive to the randomized data regression in the overlap region being an accurate reflection of the truth, as highlighted in the simulation results. When the overlap region is small, the conditional mean relationships estimated from the overlap region may be misspecified, leading to bias and large variability in estimates of unmeasured confounding bias. To assess goodness of fit, investigators can compare estimates to the truth in the randomized data overlap region. Regularization and cross-validation can reduce chances of overfitting to the data, particularly with more flexible regression approaches. Further practical challenges to applying CCDS estimators in other settings may include imperfect covariate correspondence between observational and randomized data sources. 
Our approach assumes that, after incorporating common covariates, there are no unmeasured outcome determinants (for estimating PTSMs) or effect modifiers (for estimating PATEs) that differ in distribution between randomized and observational groups. However, if this assumption is violated, CCDS estimators often performed better than using randomized data alone.

	Future extensions to the CCDS estimation framework could consider addressing positivity of treatment assignment violations, combining more than two studies (with at least one randomized and one observational), alternative approaches for determining the overlap region that allow for the degree of information borrowing to depend on the similarity of randomized and observational observations, and overlap estimation that does not rely on an estimated propensity score for selection, such as a convex hull approach \citep{king2006} or estimating common causal support \citep{hill2013}. Randomized and observational data commonly face multiple challenges beyond those of positivity of selection violation and unmeasured confounding discussed here. These challenges include lack of independence between observations (e.g., clustering), missing data, and measurement error. Methods for addressing such challenges can be combined with our CCDS approaches.

	Our CCDS estimators have relevance to many other settings. Positivity of selection violation and unmeasured confounding arise in other studies where the target population is composed of randomized and observational subsets, or more broadly when observational data are being combined with randomized data. For example, in comprehensive cohort studies, patients who refuse randomization are enrolled in a parallel observational study \citep{lu2019,olschewski1985} and when randomized controlled trials are embedded in electronic health record data, the observational data can provide information on patients included in and excluded from the trial \citep{kibbelaar2017}. Policy evaluation studies can be combined with  observational data from outside the evaluation geography to estimate scale-up impacts \citep{attanasio2003, kern2016}. Across these settings, CCDS estimators can be used to generalize to the target population represented by the union of the randomized and observational data. CCDS could also be applied when randomized data represent the target population but will be combined with observational data to increase power, such as in clinical trials that use a mix of randomized and historical controls \citep{ghadessi2020}, or when, in the absence of a comprehensive target sample, a combination of randomized and observational studies may more fully represent the target population than either study alone \citep{prentice2005,vaitsiakhovich2018}.
	
	Generalizability methods applied to a specified target population are necessary to obtain unbiased estimates for a policy-relevant population. The internal validity of randomized studies is insufficient to obtain unbiased causal estimates; external validity also needs to be considered. The CCDS estimators presented here provide several approaches for combining randomized with observational data to make inferences that do not rely on extrapolating beyond randomized data support nor on the assumption of unmeasured confounding in the observational data.
	
	\begin{centering}
		\section{SUPPLEMENTARY MATERIALS}  
	\end{centering}

	\begin{description}
	
		\item[Supplemental Materials:] Derivations, proofs, implementation details, additional estimators, supplemental simulations, and further NYC Medicaid data descriptions and results. (pdf)
	
	\end{description}

	\bibliographystyle{Chicago}
	\bibliography{CCDS_for_generalizability.bib}

\end{document}


\def\spacingset#1{\renewcommand{\baselinestretch}%
		{#1}\small\normalsize} \spacingset{1}

	
	\if1\blind
	{
		\title{\bf Supplemental Material for ``Conditional Cross-Design Synthesis Estimators for Generalizability in Medicaid''}
		\author{Irina Degtiar\textsuperscript{a}\thanks{Contact: Dr. Irina Degtiar, Department of Biostatistics, Harvard T.H. Chan School of Public Health, Boston, MA. (email: \href{mailto:idegtiar@g.harvard.edu}{idegtiar@g.harvard.edu}). This work was supported by NIH New Innovator Award under Grant DP2MD012722
				and NIH under training grants T32LM012411 and T32ES07142.}, Tim Layton\textsuperscript{b}, Jacob Wallace\textsuperscript{c}, and Sherri Rose\textsuperscript{d} \hspace{.2cm}\\
			\small
			\textsuperscript{a} Harvard T.H. Chan School of Public Health\\
			\small
			\textsuperscript{b} Harvard Medical School\\
			\small
			\textsuperscript{c} Yale School of Public Health\\
			\small
			\textsuperscript{d} Stanford University}
		\maketitle
	} \fi

	\if0\blind
	{
		\bigskip
		\bigskip
		\bigskip
		\begin{center}
			{\LARGE\bf Supplemental Material for ``Conditional Cross-Design Synthesis Estimators for Generalizability in Medicaid''}
		\end{center}
		\medskip
	} \fi

	\newpage
	\addtolength{\textheight}{.5in}%
	
	\renewcommand{\thetable}{S\arabic{table}}
	\renewcommand{\thefigure}{S\arabic{figure}}
	\section{Derivation of Assumption 1b}\label{appendix:assumptionDerivation}
	
	To overcome violations of Assumptions 1 (mean conditional treatment exchangeability, or no unmeasured confounding) and 5 (positivity of study selection), we can leverage information from the combination of randomized and observational data.
	
	For unmeasured confounding bias, we begin by characterizing the conditional bias in the observational group:
	$b(a,\bm{x}) \equiv E(Y^a|S=0,A=a,\bm{X}=\bm{x})-E(Y^a|S=0,\bm{X}=\bm{x})$.
	The conditional bias corresponds to the average difference in potential outcomes between observational group individuals on intervention $a$ vs. marginally, conditioning on measured covariates $\bm{X}=\bm{x}$. We could alternatively have defined conditional bias relative to a specific alternative intervention, $E(Y^a|S=0,A=a',\bm{X}=\bm{x})$, or relative to all other interventions, $E(Y^a|S=0,A \ne a,\bm{X}=\bm{x})$; the same principles hold. Mean conditional treatment exchangeability holds if and only if $b(a,\bm{X}) = 0$ for all $a \in \mathcal{A}$. 
	
	By randomization, mean conditional treatment exchangeability holds for the randomized group, hence
	$E(Y^a|S=1,A=a,\bm{X}=\bm{x})-E(Y^a|S=1,\bm{X}=\bm{x})=0$.
	We therefore have that:
	\begin{align*}
		b(a,\bm{x})&=E(Y^a|S=0,A=a,\bm{X}=\bm{x})-E(Y^a|S=1,A=a,\bm{X}=\bm{x}) \\
		&\quad- [E(Y^a|S=0,\bm{X}=\bm{x})-E(Y^a|S=1,\bm{X}=\bm{x})].
	\end{align*}
	By mean conditional exchangeability for study selection, $E(Y^a|S=1,\bm{X})=E(Y^a|S=0,\bm{X})$ and thus
	$b(a,\bm{x})=E(Y^a|S=0,A=a,\bm{X}=\bm{x})-E(Y^a|S=1,A=a,\bm{X}=\bm{x})$.
	
	However, overlapping support between randomized and observational groups only exists in $\mathcal{R}_\text{overlap}$, hence $E(Y^a|S=0,A=a,\bm{X}=\bm{x})-E(Y^a|S=1,A=a,\bm{X}=\bm{x})$ can only be identified in $\mathcal{R}_\text{overlap}$ without further assumptions to warrant the extrapolation. One extrapolation approach would be to directly extrapolate from the randomized group to obtain potential outcomes in regions of non-support (see the rand estimator in Section \ref{sec:comparison_estimators} of the main manuscript for an estimation strategy based on this approach), or to extrapolate for the purposes of estimating bias in regions of non-support (see the 2-stage whole data estimator in Appendix \ref{appendix:2stageEstimator} of this document), but estimation relying on these strategies is sensitive to parametric assumptions needed to extrapolate beyond the randomized data's support. We instead make an alternative assumption, Assumption 1b: $b(a,\bm{x})=b(a,\bm{x}|R_\text{ overlap}=1)$; namely, that the same conditional bias relationship that holds in the region of overlap also holds in the broader support of the observational group. When estimating PTSMs, more weakly, the constant conditional bias assumption must hold in expectation over the $\bm{X}$ distribution in the observational data: $E_{\bm{X}}\big[ b(a,\bm{x}) \big| S=0 \big] = E_{\bm{X}}\big[b(a,\bm{x}|R_\text{ overlap}=1) \big| S=0 \big]$.
	
	The mean constant conditional bias assumption can also be restated with respect to the unmeasured confounders that are implicitly being integrated over. Assumption 1b states that, in expectation, the bias when integrating over the distribution of unmeasured confounders in $\mathcal{R}_\text{overlap}$ is equivalent to the bias when integrating over the distribution of unmeasured confounders in $\mathcal{R}_\text{obs}$. Namely, with $\bm{U}$ corresponding to unmeasured confounders, the mean constant bias assumption can be written as: 
	\begin{align*}
		E_{\bm{X}}\Big\{E_{\bm{U}}\big[&E(Y^{a}|S=0,A=a,R_\text{overlap}=1,\bm{X},\bm{U})|S=0,A=a,R_\text{overlap}=1,\bm{X}\big] \\ &-E(Y^{a}|S=1,A=a,R_\text{overlap}=1,\bm{X})\Big|S=0\Big\} \\ = 
		E_{\bm{X}}\Big\{E_{\bm{U}}\big[&E(Y^{a}|S=0,A=a,\bm{X},\bm{U})|S=0,A=a,\bm{X}\big] \\ &-E(Y^{a}|S=1,A=a,\bm{X})\Big|S=0\Big\}
	\end{align*} for all $a \in \mathcal{A}$.
	
	\section{Sensitivity Analysis Bounds}\label{appendix:sensAnalysis}
	Making no constant conditional bias assumptions, we arrive at the following functional of the observed data and potential outcomes:
	
	\resizebox{\hsize}{!}{
		\begin{minipage}{1.1\linewidth}
			\begin{align}
				E(Y^{a}) =& E_{\bm{X}}[E(Y|S=1,A=a,\bm{X})|S=1]P(S=1) \nonumber\\
				&+  E_{\bm{X}}[E(Y|S=0,A=a,\bm{X})-b'(a,x)P(A \ne a|S=0,\bm{X})|S=0]P(S=0), \label{eq:noassumptionsbias}
			\end{align}
			where $b'(a,\bm{x})=E(Y^a|S=0,A=a,\bm{X}=\bm{x})-E(Y^a|S=0,A \ne a,\bm{X}=\bm{x}).\\$
		\end{minipage}
	}
	
	\emph{Proof for \ref{eq:noassumptionsbias}:}
	Using the law of iterated expectations, no unmeasured confounding in the randomized group, SUTVA assumptions, and positivity assumptions, we obtain the following:
	
	\resizebox{\hsize}{!}{
		\begin{minipage}{1.1\linewidth}
			\begin{align*}
				E(Y^{a}) &= E(Y^{a}|S=1)P(S=1) + E(Y^{a}|S=0)P(S=0) \\
				&= E_{\bm{X}}[E(Y^{a}|S=1,A=a,\bm{X})|S=1]P(S=1) \nonumber\\
				&\qquad +  E_{\bm{X}}[E(Y^{a}|S=0,A=a,\bm{X})P(A=a|S=0,\bm{X}) \nonumber\\
				&\qquad +E(Y^{a}|S=0,A \ne a,\bm{X})P(A \ne a|S=0,\bm{X})|S=0]P(S=0) \\
				&= E_{\bm{X}}[E(Y^{a}|S=1,A=a,\bm{X})|S=1]P(S=1) \nonumber\\
				&\qquad +  E_{\bm{X}}[E(Y^{a}|S=0,A=a,\bm{X})P(A=a|S=0,\bm{X}) \nonumber\\
				&\qquad +\{E(Y^{a}|S=0,A=a,\bm{X})-b'(a,x)\}P(A \ne a|S=0,\bm{X})|S=0]P(S=0) \\
				&= E_{\bm{X}}[E(Y^{a}|S=1,A=a,\bm{X})|S=1]P(S=1) \nonumber\\
				&\qquad +  E_{\bm{X}}[E(Y^{a}|S=0,A=a,\bm{X})-b'(a,x)P(A \ne a|S=0,\bm{X})|S=0]P(S=0) \\
				&= E_{\bm{X}}[E(Y|S=1,A=a,\bm{X})|S=1]P(S=1) \nonumber\\
				&\qquad +  E_{\bm{X}}[E(Y|S=0,A=a,\bm{X})-b'(a,x)P(A \ne a|S=0,\bm{X})|S=0]P(S=0).
			\end{align*}
			\linebreak
		\end{minipage}
	}
	
	Identity \eqref{eq:noassumptionsbias} can be used as the basis for sensitivity analysis, substituting different plausible bias relationships for $b'(a,\bm{x})$, as was done by \cite{brumback2004}. In many settings, it is unlikely that $b'(a,\bm{x})$ would have different signs for different $a$ (this would imply that within the same level of $\bm{X}$, individuals would have the largest outcome on the treatment they ended up on compared to other treatments). Among the various possible functional forms for the bias term presented in \cite{brumback2004}, we could assume bias would depend on measured covariates and take the form $b'(a,\bm{x})=\bm{\beta_a} \bm{X}$. Note the similarity to the 2-stage conditional cross-design synthesis (CCDS) approach where a slightly different formulation of the bias term is estimated from the overlap region.  
	
	Identity \eqref{eq:noassumptionsbias} highlights that the bias from the naive obs/rand estimator in Section \ref{sec:comparison_estimators} that averages across randomized and observational estimates for randomized and observational units respectively is therefore:
	\begin{align*}
		E&_{\bm{X}}\big[b'(a,x)P(A \ne a|S=0,\bm{X}) \big| S=0\big]P(S=0) \nonumber\\
		&= E_{\bm{X}}\Big[\big\{E(Y^a|S=0,A=a,\bm{X}=\bm{x})-E(Y^a|S=0,A \ne a,\bm{X}=\bm{x})\big\} \nonumber\\ & \quad \times P(A \ne a|S=0,\bm{X}) \big| S=0\Big]P(S=0).
	\end{align*} 
	
	\section{Proof for Identification of $\psi_\text{CCDS}(a)$}\label{appendix:identificationProof}
	\setcounter{equation}{0} 
	
	\resizebox{\hsize}{!}{
		\begin{minipage}{1.1\linewidth}
			\begin{align}
				E(Y^{a}) &= E(Y^{a}|S=1)P(S=1) + E(Y^{a}|S=0)P(S=0) \label{eq:proof_psi_line1} \\
				&= E_{\bm{X}}[E(Y^{a}|S=1,\bm{X})|S=1]P(S=1) + E_{\bm{X}}[E(Y^{a}|S=0,\bm{X})|S=0]P(S=0) \label{eq:proof_psi_line2} \\
				&= E_{\bm{X}}[E(Y^{a}|S=1,\bm{X})|S=1]P(S=1) + E_{\bm{X}}[E(Y^{a}|S=1,\bm{X})|S=0]P(S=0) \label{eq:proof_psi_line3}\\
				&= E_{\bm{X}}[E(Y^{a}|S=1,A=a,\bm{X})|S=1]P(S=1) \nonumber\\ &\quad + E_{\bm{X}}[E(Y^{a}|S=1,A=a,\bm{X})|S=0]P(S=0) \label{eq:proof_psi_line4}\\
				&= E_{\bm{X}|S=1}[E(Y^{a}|S=1,A=a,\bm{X}|S=1)]P(S=1) + E_{\bm{X}}[E(Y^{a}|S=0,A=a,\bm{X}) \nonumber\\ &\quad - \{E(Y^{a}|S=0,A=a,\bm{X}) - E(Y^{a}|S=1,A=a,\bm{X})\}|S=0]P(S=0) \label{eq:proof_psi_line5}\\
				&= E_{\bm{X}}[E(Y^{a}|S=1,A=a,\bm{X})|S=1]P(S=1) + E_{\bm{X}}[E(Y^{a}|S=0,A=a,\bm{X}) \nonumber\\ &\quad - \{E(Y^{a}|S=0,A=a,R_\text{overlap}=1,\bm{X}) - E(Y^{a}|S=1,A=a,R_\text{overlap}=1,\bm{X})\}|S=0]P(S=0) \label{eq:proof_psi_line6}\\
				&= E_{\bm{X}}[\underbrace{E(Y|S=1,A=a,\bm{X})|S=1]}_\text{(a) RCT contribution}P(S=1) + E_{\bm{X}}[\underbrace{E(Y|S=0,A=a,\bm{X})}_{\substack{\text{(b) preliminary} \\ \text{observational contribution}}}\nonumber\\ &\quad -  \underbrace{\{E(Y|S=0,A=a,R_\text{overlap}=1,\bm{X}) - E(Y|S=1,A=a,R_\text{overlap}=1,\bm{X})\}}_\text{(c) debiasing term for observational contribution}|S=0]P(S=0). \label{eq:proof_psi_line7}
			\end{align}
		\end{minipage}
	}

	Lines \eqref{eq:proof_psi_line1} and \eqref{eq:proof_psi_line2} follow from the law of iterated expectations. Line \eqref{eq:proof_psi_line3} follows from Assumption 4 of conditional exchangeability for study selection; line \eqref{eq:proof_psi_line4} follows from the first part of Assumption 1b: $E(Y^{a}|S=1,A=a,\bm{X})=E(Y^{a}|S=1,\bm{X})$; line \eqref{eq:proof_psi_line5} adds and subtracts the same term; line \eqref{eq:proof_psi_line6} then follows from the constant conditional bias part of Assumption 1b; line \eqref{eq:proof_psi_line7} follows from Assumptions 3 and 6 of SUTVA for treatment assignment and study selection; the final quantities are well-defined by the two positivity assumptions, 2 and 5.
	
	One can alternatively identify treatment-specific means through different decompositions of the data in lines \eqref{eq:proof_psi_line2}-\eqref{eq:proof_psi_line3} (see Appendix \ref{appendix:altDecomp}). Each functional implies different estimation strategies that rely on different auxiliary regression models. 
	
	\section{Estimators From Alternative Decompositions}\label{appendix:altDecomp}
	\setcounter{equation}{0} 
	One can identify target population treatment-specific means (PTSMs) through functionals derived from alternative decompositions of the target population probability distribution. We present estimators $\hat{\psi}_2(a)$ and $\hat{\psi}_3(a)$ from two such alternative decompositions. Identification of $\psi_2(a)$ and $\psi_3(a)$ relies on a slightly different formulation of Assumption 1b: $b(a,x|R_\text{obs-only}=1)=b(a,x|R_\text{ overlap}=1)$, i.e.,
	\begin{align*}
		E(Y^{a}|S=0,A=a,R_\text{overlap}=1,\bm{X})&-E(Y^{a}|S=1,A=a,R_\text{overlap}=1,\bm{X}) =\\ 
		E(Y^{a}|S=0,A=a,R_\text{obs-only}=1,\bm{X})&-E(Y^{a}|S=1,A=a,R_\text{obs-only}=1,\bm{X}).
	\end{align*}
	
	Under this alternative formulation of Assumption 1b, along with Assumptions 2 - 5b, we can identify the causal estimand as follows:
	
		 $\begin{aligned}[t] 
			\psi_2(a)=&E_{\bm{X}}\Big[E(Y|S=1,A=a,\bm{X})\Big|  R_\text{RCT}=1\Big] P(R_\text{RCT}=1) \\
			& + E_{\bm{X}}\Big[E(Y|S=0,A=a,R_\text{obs-only}=1,\bm{X}) - \big\{E(Y|S=0,A=a,R_\text{overlap}=1,\bm{X}) \\
			&\qquad \qquad - E(Y|S=1,A=a,R_\text{overlap}=1,\bm{X})\big\}\Big|R_\text{obs-only}=1\Big]P(R_\text{obs-only}=1),   
		\end{aligned}$
		
		 $\begin{aligned}[t] 
			\psi_3(a)=&E_{\bm{X}}\Big[E(Y|S=1,A=a,\bm{X})\Big|S=1\Big] P(S=1) \\
			& + E_{\bm{X}}\Big[E(Y|S=1,A=a,R_\text{overlap}=1,\bm{X})\Big|S=0,R_\text{overlap}=1\Big]  P(S=0, R_\text{overlap}=1) \\
			& + E_{\bm{X}}\Big[E(Y|S=0,A=a,R_\text{obs-only}=1,\bm{X}) - \big\{E(Y|S=0,A=a,R_\text{overlap}=1,\bm{X}) \\
			&\qquad - E(Y|S=1,A=a,R_\text{overlap},\bm{X})=1\big\}\Big|R_\text{obs-only}=1\Big]P(R_\text{obs-only}=1).   
		\end{aligned}$

	\pagebreak
	\emph{Proof for $\psi_2(a)$:}
	
	\resizebox{\hsize}{!}{
		\begin{minipage}{1.1\linewidth}
			\begin{align}
				E(Y^{a}) &= E(Y^{a}|R_\text{RCT}=1)P(R_\text{RCT}=1) + E(Y^{a}|R_\text{obs-only}=1)P(R_\text{obs-only}=1) \label{eq:proof_psi2_line1} \\
				&= E_{\bm{X}}[E(Y^{a}|R_\text{RCT}=1,\bm{X})|R_\text{RCT}=1]P(R_\text{RCT}=1) \nonumber\\ &\quad + E_{\bm{X}}[E(Y^{a}|R_\text{obs-only}=1,\bm{X})|R_\text{obs-only}=1]P(R_\text{obs-only}=1) \label{eq:proof_psi2_line2} \\
				&= E_{\bm{X}}[E(Y^{a}|S=1,R_\text{RCT}=1,\bm{X})|R_\text{RCT}=1]P(R_\text{RCT}=1) \nonumber\\ &\quad + E_{\bm{X}}[E(Y^{a}|S=1,R_\text{obs-only}=1,\bm{X})|R_\text{obs-only}=1]P(R_\text{obs-only}=1) \label{eq:proof_psi2_line3} \\
				&= E_{\bm{X}}[E(Y^{a}|S=1,A=a,\bm{X})|R_\text{RCT}=1]P(R_\text{RCT}=1) \nonumber\\ &\quad + E_{\bm{X}}[E(Y^{a}|S=1,A=a,R_\text{obs-only}=1,\bm{X})|R_\text{obs-only}=1]P(R_\text{obs-only}=1) \label{eq:proof_psi2_line4} \\
				&= E_{\bm{X}}[E(Y^{a}|S=1,A=a,\bm{X})|R_\text{RCT}=1]P(R_\text{RCT}=1) \nonumber\\ &\quad + E_{\bm{X}}[E(Y^{a}|S=0,A=a,R_\text{obs-only}=1,\bm{X}) \nonumber\\ &\quad - \{E(Y^{a}|S=0,A=a,R_\text{obs-only}=1,\bm{X}) \nonumber\\ &\quad - E(Y^{a}|S=1,A=a,R_\text{obs-only}=1,\bm{X})\}|R_\text{obs-only}=1]P(R_\text{obs-only}=1) \label{eq:proof_psi2_line5} \\
				&= E_{\bm{X}}[E(Y^{a}|S=1,A=a,\bm{X})|R_\text{RCT}=1]P(R_\text{RCT}=1) \nonumber\\ &\quad + E_{\bm{X}}[E(Y^{a}|S=0,A=a,R_\text{obs-only}=1,\bm{X}) \nonumber\\ &\qquad -\{E(Y^{a}|S=0,A=a,R_\text{overlap}=1,\bm{X}) \nonumber\\ &\qquad - E(Y^{a}|S=1,A=a,R_\text{overlap}=1,\bm{X})\}|R_\text{obs-only}=1]P(R_\text{obs-only}=1) \label{eq:proof_psi2_line6} \\
				&= \underbrace{E_{\bm{X}}[E(Y|S=1,A=a,\bm{X})|R_\text{RCT}=1]}_{\substack{\text{(a) RCT contribution and observational } \\ \text{contribution in region of overlap}}}P(R_\text{RCT}=1) + E_{\bm{X}}[\underbrace{E(Y|S=0,A=a,R_\text{obs-only}=1,\bm{X})}_{\substack{\text{(b) preliminary observational contribution in } \\ \text{region of no overlap}}} \nonumber\\ &\qquad - \underbrace{\{E(Y|S=0,A=a,R_\text{overlap}=1,\bm{X}) - E(Y|S=1,A=a,R_\text{overlap}=1,\bm{X})\}}_{\substack{\text{(c) debiasing term for observational contribution in } \\ \text{region of no overlap}}}|R_\text{obs-only}=1] \times P(R_\text{obs-only}=1). \label{eq:proof_psi2_line7}
			\end{align}
		\end{minipage}
	}
	
	\pagebreak
	\emph{Proof for $\psi_3(a)$:}
	
	\resizebox{\hsize}{!}{
		\begin{minipage}{1.1\linewidth}
			\begin{align}
				E(Y^{a}) &= E(Y^{a}|S=1)P(S=1) + E(Y^{a}|S=0,R_\text{overlap}=1)P(S=0,R_\text{overlap}=1)  \nonumber\\ &\quad + E(Y^{a}|R_\text{obs-only}=1)P(R_\text{obs-only}=1) \label{eq:proof_psi3_line1} \\
				&= E_{\bm{X}}[E(Y^{a}|S=1,\bm{X})|S=1]P(S=1)  \nonumber\\ &\quad + E_{\bm{X}}[E(Y^{a}|S=1,R_\text{overlap}=1,\bm{X})|S=0,R_\text{overlap}=1]P(S=0,R_\text{overlap}=1)   \nonumber\\ &\quad + E_{\bm{X}}[E(Y^{a}|R_\text{obs-only}=1,\bm{X})|R_\text{obs-only}=1]P(R_\text{obs-only}=1) \label{eq:proof_psi3_line2} \\		
				&= \underbrace{E_{\bm{X}}[E(Y|S=1,A=a,\bm{X})|S=1]}_\text{(a) RCT contribution}P(S=1)  \nonumber\\ &\quad + \underbrace{E_{\bm{X}}[E(Y|S=1,A=a,R_\text{overlap}=1,\bm{X})|S=0,R_\text{overlap}=1]}_\text{(a) observational contribution in region of overlap}P(S=0,R_\text{overlap}=1)  \nonumber\\ &\quad +   E_{\bm{X}}[\underbrace{E(Y|S=0,A=a,R_\text{obs-only}=1,\bm{X})}_{\substack{\text{(b) preliminary observational contribution in } \\ \text{region of no overlap}}}\nonumber\\ &\qquad -  \underbrace{\{E(Y|S=0,A=a,R_\text{overlap}=1,\bm{X}) - E(Y|S=1,A=a,R_\text{overlap}=1,\bm{X})\}|R_\text{obs-only}=1}_\text{(c) debiasing term for observational contribution in region of no overlap}] \times P(R_\text{obs-only}=1). \label{eq:proof_psi3_line3}
			\end{align}
		\end{minipage}
	}
	
	As in the proof for $\psi_\text{CCDS}(a)$, lines \eqref{eq:proof_psi2_line1}, \eqref{eq:proof_psi2_line2}, and \eqref{eq:proof_psi3_line1} follow from the law of iterated expectations. Line \eqref{eq:proof_psi2_line3} follows from Assumption 4 of conditional exchangeability for study selection; line \eqref{eq:proof_psi2_line4} follows from the first part of Assumption 1b: $E(Y^{a}|S=1,A=a,\bm{X})=E(Y^{a}|S=1,\bm{X})$ (and the redundancy of $S=1$ and $\mathcal{R}_\text{RCT}$); line \eqref{eq:proof_psi2_line5} adds and subtracts the same term; line \eqref{eq:proof_psi2_line6} then follows from the constant conditional bias part of Assumption 1b; line \eqref{eq:proof_psi2_line7} follows from Assumptions 3 and 6 of SUTVA for treatment assignment and study selection; the final quantities are well-defined by the two positivity assumptions, 2 and 5. For lines \eqref{eq:proof_psi3_line2}-\eqref{eq:proof_psi3_line3}, the same steps seen in the proof of $\psi_2(a)$ were repeated to arrive at the final functional.
	
	Each of these three functionals ($\psi_\text{CCDS}, \psi_2, \psi_3$) suggest slightly different estimation procedures that rely on different auxiliary regression models for different subsets of data. For example, the outcome regression estimators of $\psi_2$ and $\psi_3$ would be as follows:	 
	
	\resizebox{\hsize}{!}{
		\begin{minipage}{1.1\linewidth}
			\begin{align*}
				&\hat{\psi}_{2\text{-OR}}(a)& = &\frac{1}{n}\sum_{i=1}^{n} \underbrace{\hat{Q}_i(S_i=1,A_i=a,\bm{X}_i)\mathbbm{1}( R_\text{RCT}=1)}_{\substack{\text{(a) RCT estimate and observational } \\ \text{estimate in region of overlap}}} +  \underbrace{\hat{Q}_i(S_i=0,A_i=a,R_\text{obs-only}=1,\bm{X}_i)\mathbbm{1}(R_\text{i, obs-only}=1)}_{\substack{\text{(b) preliminary observational estimate in } \\ \text{region of no overlap}}}  \nonumber\\ &&&\quad \quad - \underbrace{\Big\{\hat{Q}_i(S_i=0,A_i=a,R_\text{overlap}=1,\bm{X}_i) - \hat{Q}_i(S_i=1,A_i=a,R_\text{overlap}=1,\bm{X}_i)\Big\}\mathbbm{1}(R_\text{i, obs-only}=1)}_\text{(c) debiasing term for observational estimate in region of no overlap}, \\
				&\hat{\psi}_{3\text{-OR}}(a)& = &\frac{1}{n}\sum_{i=1}^{n} \underbrace{\hat{Q}_i(S_i=1,A_i=a,\bm{X}_i) \mathbbm{1}(S_i=1)}_\text{(a) RCT estimate } + 
				\underbrace{\hat{Q}_i(S_i=1,A_i=a,R_\text{overlap}=1,\bm{X}_i) \mathbbm{1}(S_i=0,R_\text{i, overlap}=1)}_{\substack{\text{(b1) preliminary observational estimate in } \\ \text{region of overlap}}} \nonumber\\ &&&\quad \quad +  \underbrace{\hat{Q}_i(S_i=0,A_i=a,R_\text{obs-only}=1,\bm{X}_i) \mathbbm{1}(R_\text{i, obs-only}=1)}_{\substack{\text{(b2) preliminary observational estimate in } \\ \text{region of no overlap}}}  \nonumber\\ &&&\quad \quad - \underbrace{\Big\{\hat{Q}_i(S_i=0,A_i=a,R_\text{overlap}=1,\bm{X}_i) - \hat{Q}_i(S_i=1,A_i=a,R_\text{overlap}=1,\bm{X}_i)\Big\}\mathbbm{1}(R_\text{i, obs-only}=1)}_\text{(c) debiasing term for observational estimate in region of no overlap}.
			\end{align*}
		\end{minipage}
	}
	
	Choices between the two estimators here and the one presented in the main paper should rely on such considerations as efficiency and which regressions may better fit  the data (e.g., both $\hat{\psi}_{2\text{-OR}}(a)$ and $\hat{\psi}_{3\text{-OR}}(a)$ rely on preliminary observational estimates estimated from regressions fit to small subsets of the data). As a reminder, $\psi_\text{CCDS}(a)$ suggests an estimation procedure in which regressions are fit using: (a) all the randomized data to estimate potential outcomes in the randomized data, (b) all the observational data to estimate preliminary potential outcomes in the observational data, and (c) the randomized data in the overlap region and the observational data in the overlap region to estimate the debiasing term for preliminary observational data estimates.
	
	In contrast, $\psi_2(a)$ suggests an estimation procedure in which regressions are fit using: (a) all the randomized data to estimate potential outcomes in the randomized and in the overlap region of the observational study, (b) the observational data in the nonoverlap region to estimate preliminary potential outcomes in the nonoverlap region of the observational study, and (c) the randomized data in the overlap region and the observational data in the overlap region to estimate the debiasing term.
	
	Correspondingly, $\psi_3(a)$ suggests an estimation procedure in which regressions are fit using: (a) all the randomized data to estimate potential outcomes in the randomized population, (b1) the randomized data in the overlap region to estimate potential outcomes in the overlap region of the observational study, (b2) the observational data in the nonoverlap region to estimate preliminary potential outcomes in the nonoverlap region of the observational study, and (c) the randomized data in the overlap region and the observational data in the overlap region to estimate the debiasing term.
	
	The three estimators differ in the flexibility of their regression specifications: the latter estimators let the covariate-outcome relationship differ in the overlap vs. nonoverlap regions. However, this flexibility comes at the cost of less information borrowing across the entire covariate distribution.

	\section{Implementation}\label{appendix:implementation}
	\subsection{CCDS-OR}
	Each of the outcome regressions in $\hat{\psi}_\text{CCDS-OR}(a)$ must appropriately capture treatment effect heterogeneity such as through including all relevant interaction terms in a least squares regression  or by using flexible nonparametric approaches that discover effect heterogeneity in a data-driven fashion, such as machine learning algorithms (keeping in mind that many such approaches do not have convergence rates that result in $\sqrt{n}$-consistency). When fitting more complex algorithms for the outcome regressions, there is a potential for overfitting to the trends in the overlap region when estimating the debiasing term (third term in $\hat{\psi}_\text{CCDS-OR}(a)$), even with regularization and cross-validation. 
	
	The CCDS framework can also be used to estimate conditional PTSMs  for the CCDS-OR and 2-stage CCDS-OR estimators via a weighted average of randomized and debiased observational conditional means, weighted by the relative proportion of randomized and observational individuals in the target population. The CCDS-OR conditional PTSM estimator is as follows: 
	\begin{align*}
		\hat{\psi}_\text{CCDS-OR}(a,\bm{x}) = &\underbrace{\frac{n_\text{rand}}{n}\hat{Q}(S=1,A=a,\bm{X}=\bm{x})}_\text{(a) RCT estimate } +  \underbrace{\frac{n_\text{obs}}{n}\hat{Q}(S=0,A=a,\bm{X}=\bm{x})}_\text{(b) preliminary observational estimate} \nonumber\\ &\quad \quad 
		- \frac{n_\text{obs}}{n} \Big\{\hat{Q}(S=0,A=a,\hat{R}_\text{overlap}=1,\bm{X}=\bm{x}) \nonumber\\ &\qquad \quad \underbrace{ \quad- \hat{Q}(S=1,A=a,\hat{R}_\text{overlap}=1,\bm{X}=\bm{x}) \Big\}}_\text{(c) debiasing term for observational estimate}.
	\end{align*}
	Rather than simply using $n_\text{study}/n$, these weights can also be replaced by sampling weights.

	\subsection{2-stage CCDS}
	In the second stage of the 2-stage CCDS estimator, a simple $\hat{g}()$ function such as $\hat{g}(\bm{X})=\bm{X}^T\hat{\theta}$ can prevent overfitting to the overlap region and thus provide added stability for estimating bias, particularly when fitting more complex $\hat{Q}(S,A,R,\bm{X})$ regressions in the first stage. Substantive knowledge can also inform choice of the $\hat{g}()$ function, such as knowledge of which measured covariates can serve as proxies for unmeasured confounders.
	
	In studies with fewer treatment groups and thus potentially more data in each one, it may be beneficial to subset to randomized overlap region data in a given treatment group for bias estimation to make sure to fully capture treatment effect heterogeneity (though this approach precludes borrowing strength across treatment groups). Step 2 of the 2-stage CCDS estimator then becomes: 
	\begin{itemize} 
		\item[(2)] $\hat{b}'(S_i=1,a,\bm{X}_i)= \frac{\hat{w}_\text{bias}(S_i,A_i,\bm{X}_i)}{\sum_{i=1}^n\hat{w}_\text{bias}(S_i,A_i,\bm{X}_i)} \hat{g}(\bm{X}_i)$ with \newline $\hat{w}_\text{bias}(S_i,A_i,\bm{X}_i)=\frac{\mathbbm{1}(S_i=1,A_i=a,R_\text{overlap, i}=1)\hat{P}(S_i=0|\bm{X}_i)}{\hat{P}(R_\text{overlap, i}=1|S_i=1,\bm{X}_i)\hat{P}(S_i=1|\bm{X}_i)\hat{P}(A_i=a|S_i=1,R_\text{overlap, i}=1,\bm{X}_i)}$.
	\end{itemize}
	
	\subsection{CCDS-IPW}
	To circumvent unstable weights for CCDS-IPW and other novel estimators using weights, propensity scores and their products used in weight denominators can be trimmed. However, trimming weights effectively changes the estimand of interest, thus requiring a bias-variance tradeoff \citep{potter1993, crump2009, lee2011}.

	
	\section{2-stage Whole Data Outcome Regression Estimator} \label{appendix:2stageEstimator}
	\subsection{Estimator}
	An alternative to the constant conditional bias assumption is to instead extrapolate from the randomized study to regions not supported in the randomized study covariate distribution, $\mathcal{R}_\text{obs-only}$. If we believe that we can reliably extrapolate from the randomized data for the purpose of debiasing term estimation (although we are not confident enough to directly extrapolate potential outcomes), the 2-stage whole data (WD) outcome regression estimator would provide more power than the 2-stage CCDS approach by not restricting debiasing term estimation to the overlap region:
	\begin{itemize}
		\item[(1)] $\hat{b}'(S_i=1,a,\bm{X}_i)= \hat{Q}_i(S=0,A=a,\bm{X})\mathbbm{1}(S_i=1) - \hat{Q}_i(S=1,A=a,\bm{X})\mathbbm{1}(S_i=1)$
		\item[(2)] $\hat{b}'(S_i=1,a,\bm{X}_i)= \frac{\hat{w}_\text{bias}(S_i,\bm{X}_i)}{\sum_{i=1}^n\hat{w}_\text{bias}(S_i,\bm{X}_i)} \hat{g}(\bm{X}_i)$ with $\hat{w}_\text{bias}(S_i,\bm{X}_i)=\frac{\mathbbm{1}(S_i=1)\hat{P}(S_i=0|\bm{X}_i)}{\hat{P}(S_i=1|\bm{X}_i)}$.
	\end{itemize} 
	
	\noindent One could likewise subset to randomized data in a given treatment group for bias estimation. The 2-stage WD estimator then becomes: \begin{itemize}
		\item[(1)] $\hat{b}'(S_i=1,a,\bm{X}_i)= \hat{Q}_i(S=0,A=a,\bm{X})\mathbbm{1}(S_i=1,A_i=a) - \hat{Q}_i(S=1,A=a,\bm{X})\mathbbm{1}(S_i=1,A_i=a)$
		\item[(2)] $\hat{b}'(S_i=1,a,\bm{X}_i)= \frac{\hat{w}_\text{bias}(S_i,A_i,\bm{X}_i)}{\sum_{i=1}^n\hat{w}_\text{bias}(S_i,A_i,\bm{X}_i)} \hat{g}(\bm{X}_i)$ with $\hat{w}_\text{bias}(S_i,A_i,\bm{X}_i)=\frac{\mathbbm{1}(S_i=1,A_i=a)\hat{P}(S_i=0|\bm{X}_i)}{\hat{P}(S_i=1|\bm{X}_i)\hat{P}(A_i=a|S_i=1,A_i=a,\bm{X}_i)}$.
	\end{itemize}

	A similar approach was taken by \cite{kallus2018} to estimate target population conditional average treatment effects for a target population represented by the observational data, using $Y_i\mathbbm{1}(S_i=1,A_i=a)/\hat{P}(A_i=a|S_i=1,\bm{X}_i)$ and not $\hat{Q}_i(S=1,A=a,\bm{X})\mathbbm{1}(S_i=1)$ in Stage (1) and not weighting Stage (2). Therefore, the Kallus et al. 2-stage approach optimizes for mean squared error across the covariate distribution in the randomized group rather than the observational group, a covariate distribution that does not represent the one where we wish to minimize bias. However, it does not suffer from potentially increased variability due to the weights. The Kallus et al. 2-stage approach does not directly extend to estimating PTSMs.

	\subsection{Simulation Results}
\begin{figure}
		\centering
		\begin{subfigure}[t]{0.7\textwidth}
			\centering
			\includegraphics[width=\linewidth]{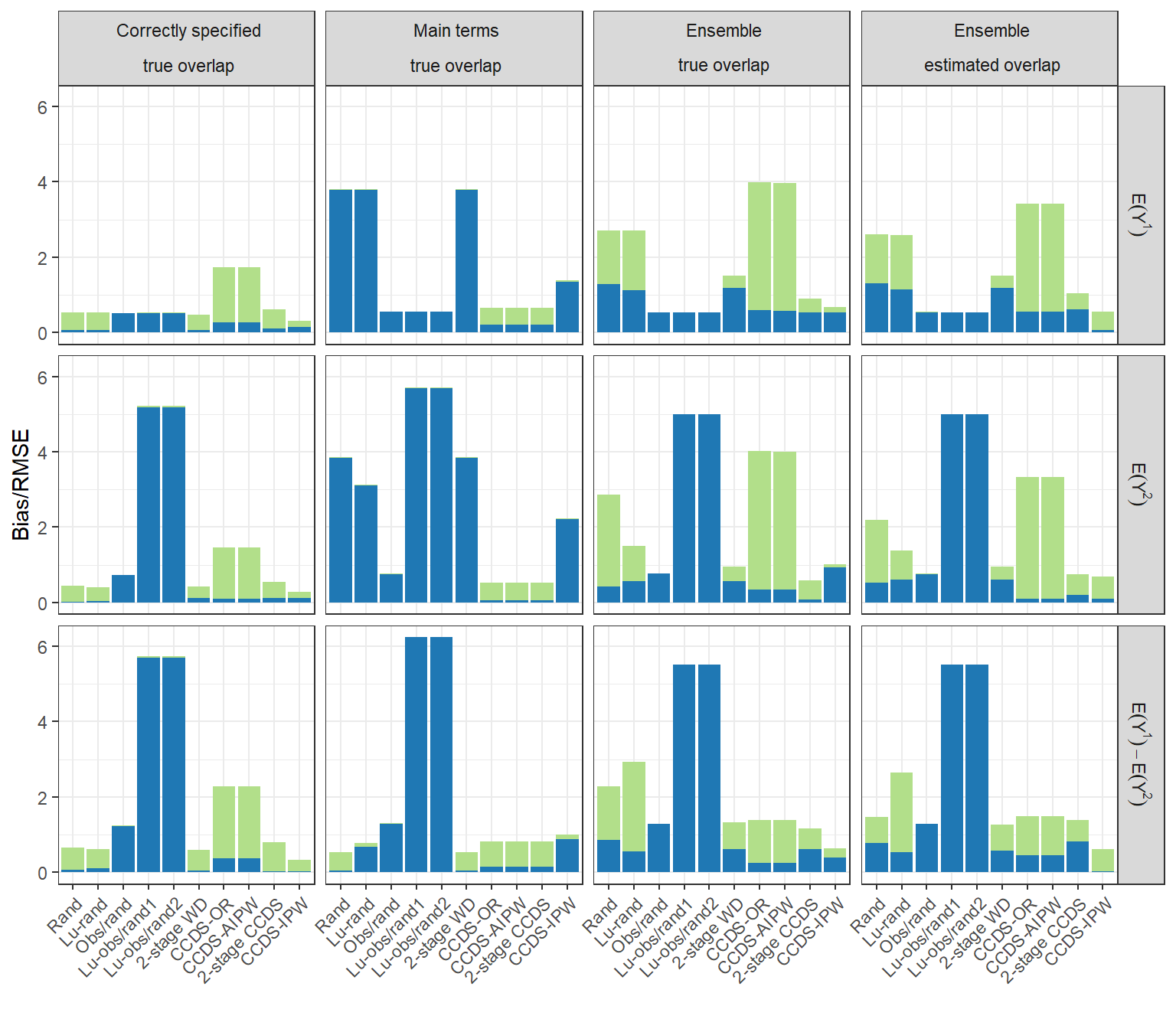} 
			\caption{Bias and RMSE} \label{fig:bias_RMSE}
		\end{subfigure}
		\begin{subfigure}[t]{0.8\textwidth}
			\centering
			\includegraphics[width=\linewidth]{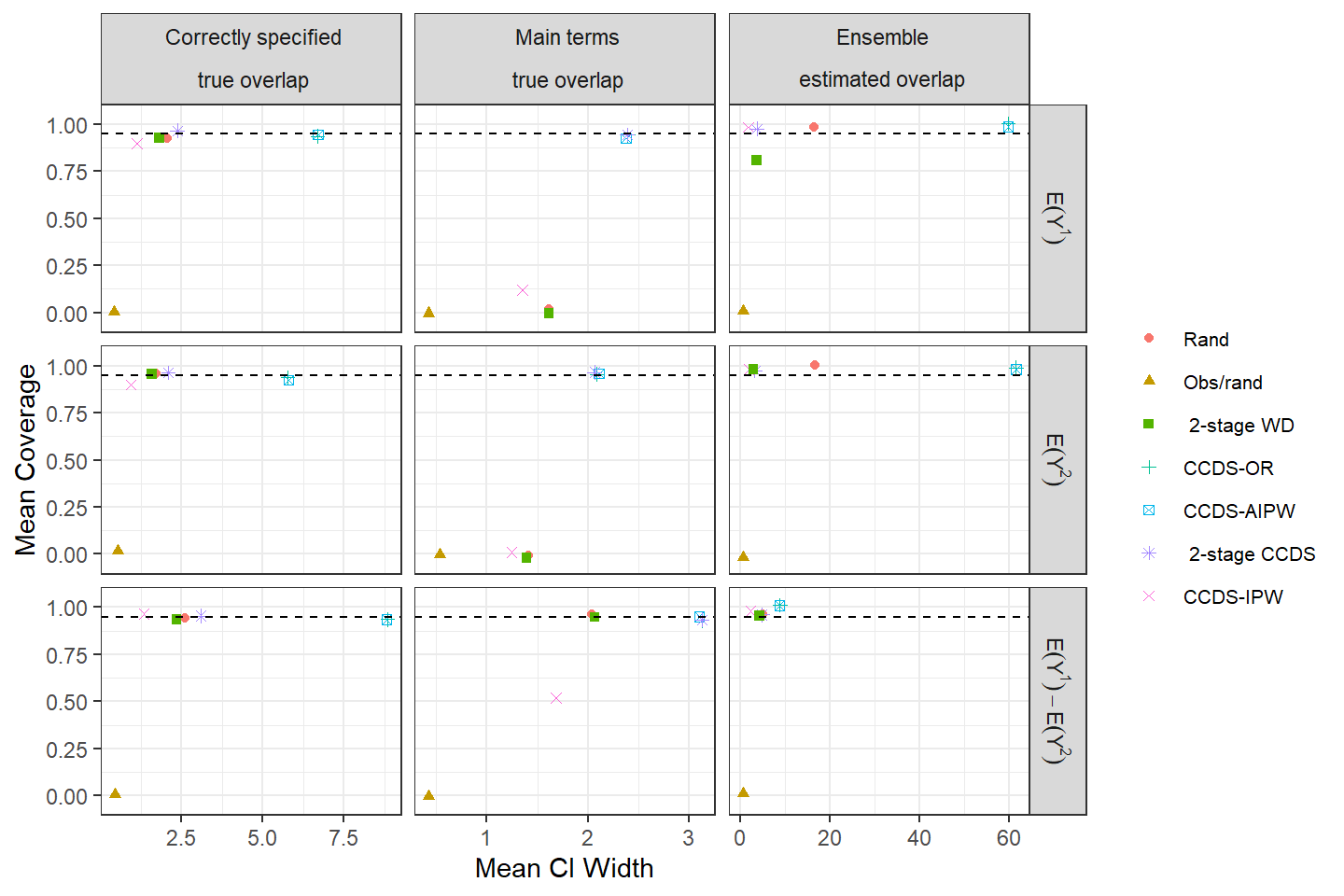} 
			\caption{Coverage and CI Width} \label{fig:coverage}
		\end{subfigure}
		\caption{Performance Across All Estimators. Panel (\subref{fig:bias_RMSE}) depicts absolute bias is the darker portion of each bar; RMSE corresponds to the total bar size. In panel (\subref{fig:coverage}), the dashed line corresponds to the target coverage of 95\%.}\label{figure:full_sim_results}
	\end{figure}

	With correctly specified regressions, all novel estimators, including the 2-stage WD approach, were able to decrease unmeasured confounding bias. The 2-stage WD approach was the most efficient novel outcome regression estimator because it used more data to fit regressions compared to CCDS estimators (Figure \ref{figure:full_sim_results}). However, when (incorrectly) fitting main terms regressions, just as with the rand estimator, extrapolation became an issue for the 2-stage WD estimator. As expected, with linear additive regressions like the correctly specified and main terms regressions, the 2-stage WD estimator is numerically equivalent to the rand estimator. Using ensemble methods the bias for the  2-stage WD estimator  tended to be similar to that of the rand estimator. Because of the 2-stage WD estimator's sensitivity to model misspecificiation, we do not generally recommend using this estimator. The estimator's poor performance highlights the importance of focusing on the overlap region for estimating unmeasured confounding bias.

	\section{Proof for $\hat{\psi}_\text{CCDS-IPW}(a)$}\label{appendix:proofCCDSIPW}
	\setcounter{equation}{0} 
	\begin{align*}
		\psi_\text{CCDS}(a) &= \underbrace{E_{\bm{X}}\Big[E(Y|S=1,A=a,\bm{X})\Big|S=1\Big]}_\text{(1)} P(S=1) \\
		&\quad + \underbrace{E_{\bm{X}}\Big[E(Y|S=0,A=a,\bm{X})|S=0\Big]}_\text{(2)}P(S=0) \\ 	
		&\quad- \big\{\underbrace{E_{\bm{X}}\Big[E(Y|S=0,A=a,R_\text{overlap}=1,\bm{X})|S=0\Big]}_\text{(3)} \\
		&\qquad \quad - \underbrace{E_{\bm{X}}\Big[E(Y|S=1,A=a,R_\text{overlap}=1,\bm{X})\Big|S=0\Big]}_\text{(4)}\big\}P(S=0).  
	\end{align*}
	
\noindent	We can then identify each of the conditional distributions via the following propensity decomposition (using conditional probability laws and positivity assumptions). For example, for component (4):
	
	\resizebox{\hsize}{!}{
		\begin{minipage}{1.1\linewidth}
			\begin{align}
				(4) &= E_{\bm{X}}\Big[E(Y|S=1,A=a,R_\text{overlap}=1,\bm{X})\big|S=0\Big] \nonumber \\ &= \frac{1}{P(S=0)}E_{\bm{X}}\Big[\mathbbm{1}(S=0)E(Y|S=1,A=a,R_\text{overlap}=1,\bm{X})\big] \nonumber\\
				&= \frac{1}{P(S=0)}E_{\bm{X}}\Bigg[P(S=0|\bm{X})\frac{E(Y\mathbbm{1}(S=1,A=a,R_\text{overlap}=1)|\bm{X})}{P(S=1,R_\text{overlap}=1|\bm{X})P(A=a|S=1,R_\text{overlap}=1,\bm{X})}\Bigg] \nonumber\\
				&= \frac{1}{P(S=0)}E_{\bm{X}}\Bigg[E\Bigg(\frac{Y\mathbbm{1}(S=1,A=a,R_\text{overlap}=1)P(S=0|\bm{X})}{P(S=1,R_\text{overlap}=1|\bm{X})P(A=a|S=1,R_\text{overlap}=1,\bm{X})}\Bigg)\Bigg]. \label{eq:line4} \nonumber\\ \nonumber
			\end{align}
		\end{minipage}
	}
	
\noindent	For weight stabilization, we can also replace $1/P(S=0)$ with $$E(w_4)^{-1} = \Bigg[E_{\bm{X}}\Bigg(\frac{\mathbbm{1}(S=1,A=a,R_\text{overlap}=1)P(S=0|\bm{X})}{P(S=1,R_\text{overlap}=1|\bm{X})P(A=a|S=1,R_\text{overlap}=1,\bm{X})}\Bigg)\Bigg]^{-1},$$ because
	\begin{align*}
		E(w_4) &= E_{\bm{X}}\Bigg(\frac{E(\mathbbm{1}(S=1,A=a,R_\text{overlap}=1)|\bm{X})P(S=0|\bm{X})}{P(S=1,R_\text{overlap}=1|\bm{X})P(A=a|S=1,R_\text{overlap}=1,\bm{X})}\Bigg) \\&=E_{\bm{X}}\big(P(S=0|\bm{X})\big) = P(S=0).
	\end{align*}
	
	\noindent This weight stabilization creates more stability for estimation and ensure estimates are in the support of the outcome variable. 
	
	We can similarly identify each of the conditional distributions in (1) - (3) through the following propensity score decompositions:
	
	\resizebox{\hsize}{!}{
		\begin{minipage}{1.1\linewidth}
			\begin{align}
				(1) &= E(w_1)^{-1} E_{\bm{X}}\Bigg[E\Bigg(\frac{Y\mathbbm{1}(S=1,A=a)P(S=1|\bm{X})}{P(S=1|\bm{X})P(A=a|S=1,\bm{X})}\Bigg)\Bigg] \nonumber\\
				&= E(w_1)^{-1} E_{\bm{X}}\Bigg[E\Bigg(\frac{Y\mathbbm{1}(S=1,A=a)}{P(A=a|S=1,\bm{X})}\Bigg)\Bigg], \nonumber\\
				(2) &= E(w_2)^{-1} E_{\bm{X}}\Bigg[E\Bigg(\frac{Y\mathbbm{1}(S=0,A=a)P(S=0|\bm{X})}{P(S=0|\bm{X})P(A=a|S=0,\bm{X})}\Bigg)\Bigg] \nonumber\\
				&= E(w_2)^{-1} E_{\bm{X}}\Bigg[E\Bigg(\frac{Y\mathbbm{1}(S=0,A=a)}{P(A=a|S=0,\bm{X})}\Bigg)\Bigg],\nonumber\\
				(3) &= E(w_3)^{-1} E_{\bm{X}}\Bigg[E\Bigg(\frac{Y\mathbbm{1}(S=0,A=a,R_\text{overlap}=1)P(S=0|\bm{X})}{P(S=0,R_\text{overlap}=1|\bm{X})P(A=a|S=0,R_\text{overlap}=1,\bm{X})}\Bigg)\Bigg] \nonumber\\
				&= E(w_3)^{-1} E_{\bm{X}}\Bigg[E\Bigg(\frac{Y\mathbbm{1}(S=0,A=a,R_\text{overlap}=1)P(S=0|\bm{X})}{P(S=0|\bm{X})P(R_\text{overlap}=1|S=0,\bm{X})P(A=a|S=0,R_\text{overlap}=1,\bm{X})}\Bigg)\Bigg] \nonumber\\
				&= E(w_3)^{-1} E_{\bm{X}}\Bigg[E\Bigg(\frac{Y\mathbbm{1}(S=0,A=a,R_\text{overlap}=1)}{P(R_\text{overlap}=1|S=0,\bm{X})P(A=a|S=0,R_\text{overlap}=1,\bm{X})}\Bigg)\Bigg], \nonumber
			\end{align}
			\end{minipage}
			}
			
					\noindent where\\
					\resizebox{\hsize}{!}{
		\begin{minipage}{1.1\linewidth}
	
			\begin{align*}
				w_1 &= \frac{\mathbbm{1}(S=1,A=a)}{P(A=a|S=1,\bm{X})}, \\
				w_2 &= \frac{\mathbbm{1}(S=0,A=a)}{P(A=a|S=0,\bm{X})}, \\
				w_3 &= \frac{\mathbbm{1}(S=0,A=a,R_\text{overlap}=1)}{P(R_\text{overlap}=1|S=0,\bm{X})P(A=a|S=0,R_\text{overlap}=1,\bm{X})}, \\
				w_4 &= \frac{\mathbbm{1}(S=1,A=a,R_\text{overlap}=1)[1-P(S=1|\bm{X})]}{P(S=1|\bm{X})P(R_\text{overlap}=1|S=1,\bm{X})P(A=a|S=1,R_\text{overlap}=1,\bm{X})}.
			\end{align*}
		\end{minipage}
	}
	
	\section{CCDS Influence Function}\label{appendix:IF} 
	To derive the influence function for $\psi_\text{CCDS}(a)$, we first derive the influence function for each of its four conditional means:\\
	
	\resizebox{\hsize}{!}{
		\begin{minipage}{1.1\linewidth}
			\begin{itemize}
				\item[(1)] For $\chi_1(a) = E_{\bm{X}}[E(Y|S=1,A=a,\bm{X})|S=1]$, 
				$$\chi'_1(a)=\frac{1}{P(S=1)}\Big[w_1\big\{Y-E(Y|S=1,A=a,\bm{X})\big\}+S\big\{E(Y|S=1,A=a,\bm{X})-\chi_1(a)\big\}\Big].$$ 
				\item[(2)] For $\chi_2(a) = E_{\bm{X}}[E(Y|S=0,A=a,\bm{X})|S=0]$, 
				$$\chi'_2(a)=\frac{1}{P(S=0)}\Big[w_2\big\{Y-E(Y|S=0,A=a,\bm{X})\big\}+(1-S)\big\{E(Y|S=0,A=a,\bm{X})-\chi_2(a)\big\}\Big].$$
				\item[(3)] For $\chi_3(a) = E_{\bm{X}}[E(Y|S=0,A=a,R_\text{overlap}=1,\bm{X})|S=0]$, 
				\begin{align*}\chi'_3(a)=\frac{1}{P(S=0)}\Big[w_3\big\{Y-E(Y|S=0,A=a,R_\text{overlap}=1,\bm{X})\big\} \\ +(1-S)\big\{E(Y|S=0,A=a,R_\text{overlap}=1,\bm{X})-\chi_3(a)\big\}\Big].\end{align*}
				\item[(4)] For $\chi_4(a) = E_{\bm{X}}[E(Y|S=1,A=a,R_\text{overlap}=1,\bm{X})|S=0]$, 
				\begin{align*}\chi'_4(a)=\frac{1}{P(S=0)}\Big[w_4\big\{Y-E(Y|S=1,A=a,R_\text{overlap}=1,\bm{X})\big\} \\ +(1-S)\big\{E(Y|S=1,A=a,R_\text{overlap}=1,\bm{X})-\chi_4(a)\big\}\Big],\\\end{align*}
			\end{itemize}
		\end{minipage}
	}
	
	\noindent where probabilities and expectations are taken under the true model and weights are as previously defined.

	The joint influence function will then be the reweighted (by $P(S=1)$ or $P(S=0)$) sum of the 4 conditional mean influence functions:
	\begin{align*}
		\chi'(a)&=w_1\big\{Y-E(Y|S=1,A=a,\bm{X})\big\}+S\big\{E(Y|S=1,A=a,\bm{X})-\chi_1(a)\big\} \\
		&\quad + w_2\big\{Y-E(Y|S=0,A=a,\bm{X})\big\}+(1-S)\big\{E(Y|S=0,A=a,\bm{X})-\chi_2(a)\big\} \\ 
		&\quad-w_3\big\{Y-E(Y|S=0,A=a,R_\text{overlap}=1,\bm{X})\big\} \\ &\qquad -(1-S)\big\{E(Y|S=0,A=a,R_\text{overlap}=1,\bm{X})-\chi_3(a)\big\} \\
		&\quad+w_4\big\{Y-E(Y|S=1,A=a,R_\text{overlap}=1,\bm{X})\big\} \\ &\qquad +(1-S)\big\{E(Y|S=1,A=a,R_\text{overlap}=1,\bm{X})-\chi_4(a)\big\}.
	\end{align*}

	\section{Comparison Estimator: \cite{lu2019}}\label{appendix:lu2019} 
	\cite{lu2019} develop one AIPW rand estimator and two obs/rand estimators relying on positivity of study selection and/or unmeasured confounding in the observational data. Their estimators split data into $k$ folds and, within each fold, fit separate regressions in each treatment group and study type, which may result in larger variance for rare or multinomial treatments, as in our applied study. \cite{lu2019} fit generalized additive models for each regression; for better comparability, we fit either parametric or ensemble machine learning regressions to match other estimators.
		
	\section{Supplemental Simulation Descriptions and Results}\label{appendix:simulation} 
	\subsection{Further implementation details}
	Ensemble regressions were implemented using the \texttt{SuperLearner} package \citep{polley2019} and consisted of \texttt{SL.glm}, \texttt{SL.glm.interact}, \texttt{SL.glmnet} with $\alpha=0.5$, \texttt{SL.ranger} with 300 trees and a minimum node size of 5\% of the sample being fit, \texttt{SL.nnet} with 2 hidden layers, \texttt{SL.earth}, \texttt{SL.gam}, and \texttt{SL.kernelKnn}. For primary results, we conservatively estimated the overlap region using $\alpha=1\% \times \text{range}(\text{logit}(\pi_S))$ and $\beta=1\% \times \text{min}(n_\text{obs},n_\text{rand})$; that is, at least $1\%$ of observations in a given treatment group must fall within $1\%$ intervals of the logit of the propensity score. 
	
	\subsection{Further Descriptions of the Data-Generating Mechanism}

\begin{figure}
		\spacingset{1}
		\centering
		\begin{subfigure}[t]{0.45\textwidth}
			\centering
			\includegraphics[width=\linewidth, height=7.5cm]{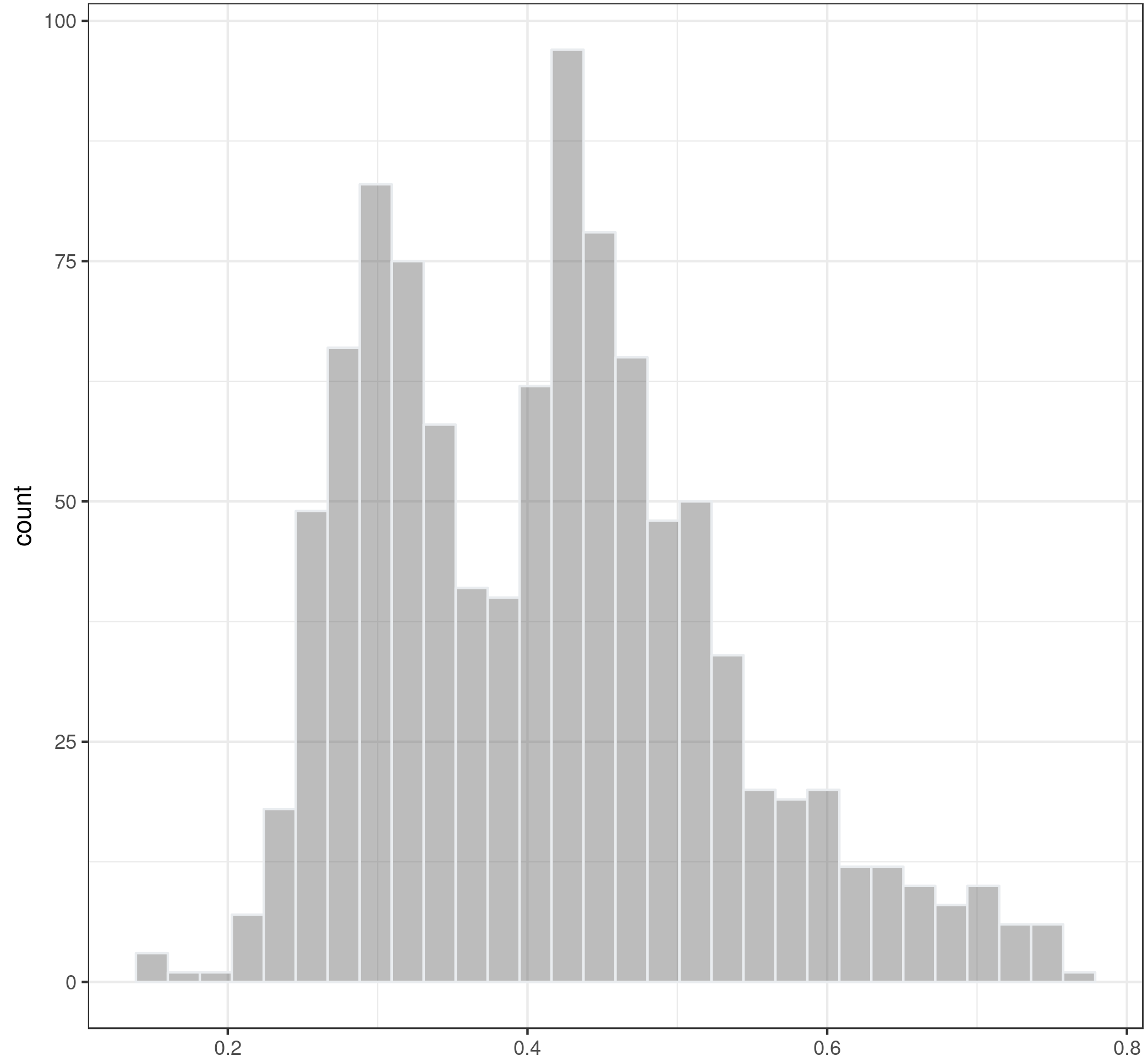}
			\caption{$\hat{P}(A=1|S=0,\bf{X},U)$}
		\end{subfigure}
		\hfill
		\begin{subfigure}[t]{0.45\textwidth}
			\centering
			\includegraphics[width=\linewidth, height=7.5cm]{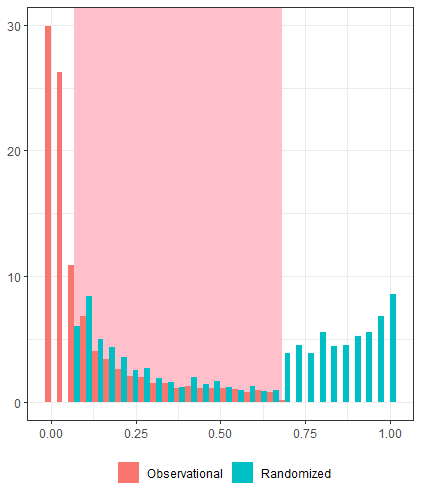}
			\caption{$\hat{P}(S=1|X)$ with true overlap region shaded} 
		\end{subfigure}
		
		\caption{Estimated Propensity Scores for Treatment and Selection.}	
		\label{figure:propensity_scores}
	\end{figure}

	The core data-generating mechanism  resulted in positivity of selection violation (Figure \ref{figure:propensity_scores}b): the confounders having varying strengths of confounding, there being relatively strong unmeasured confounding ($U$ had the second largest impact on treatment and outcome values), the conditional outcome relationship in $\mathcal{R}_\text{overlap}$ in the randomized data not fully extrapolating well to $\mathcal{R}_\text{obs-only}$ unless  the correct outcome regression was fit, and observed covariates differing in distribution across randomized and observational data. As a result, randomized and observational data each displayed external validity bias for estimating PTSMs and PATEs, and observational data likewise displayed internal validity bias due to measured and unmeasured confounding (Table \ref{table:potential_outcomes}). With these specifications, there were discrepancies between true randomized and observational study population treatment-specific means (STSMs) and study population average treatment effects (SATEs). This data-generating mechanism also ensured that identifiability assumptions held, namely:  
	\begin{enumerate}
		\item The randomized group had no unmeasured confounding and the distribution of $U$ was the same in $\mathcal{R}_\text{overlap}$ as $\mathcal{R}_\text{obs-only}$ conditioning on measured covariates; thus, the constant conditional bias assumption was satisfied (bias was in fact constant, not just conditionally constant; it was equal to $E(10U)$).
		\item Study/treatment groups had positive probabilities of receiving each treatment (Figure \ref{figure:propensity_scores}a).  
		\item Observations were independent.
		\item The unmeasured covariate did not confound the relationship between outcome and study selection. 
		\item $\mathcal{R}_\text{overlap}$ was not a null set (Figure \ref{figure:propensity_scores}).
		\item The same outcome specification held for both randomized and observational data.
	\end{enumerate}
	
	\begin{table}
		\centering
		\begin{tabular}{lccccc}
			\toprule
			& Population & \multicolumn{2}{c}{RCT} & \multicolumn{2}{c}{Obs} \\
			& Truth   & Truth      & Observed   & Truth     & Observed     \\ \midrule
			$E(Y^1)$     & 5.09       & 13.49      & 13.48        & 3.00      & 1.47         \\
			$E(Y^2)$     & 2.09       & 15.11      & 15.11        & -1.16     & 1.65         \\
			$E(Y^1-Y^2)$ & 3.00       & -1.63      & -1.62        & 4.16      & -0.19        \\ \bottomrule
		\end{tabular}
		\caption{Population and Sample True Potential Outcome Means and Means Observed in Each Treatment Group.}
		
		\label{table:potential_outcomes}
	\end{table}
	
	We additionally investigated the impacts of violations to these assumptions and of other data-generating mechanism and regression specifications on CCDS estimator performance. Expanding from the base case, we assessed the following settings: 6 regression fit specifications (main terms, squared terms, correctly specified, \texttt{ksvm}, and two ensembles), 4 target sample sizes ($n=2000, 10000, 20000, 50000$), a constant bias violation setting, 4 unmeasured confounding settings, 3 overlap settings, 3 ratios of $n_\text{RCT}$ to $n_\text{obs}$, 3 positivity of selection violation settings, 2 exchangeability of study selection violations, 5 overlap region determination settings, and 3 propensity for selection relationships. We also examined 6 alternative outcome regressions (main terms, more complex effect heterogeneity, knot, more severe knot, knot inside overlap region, and $U \times X_1$ interaction).

	\subsection{Overlap Region Specifications}
\begin{figure}
		\spacingset{1}
		\begin{center}
			\includegraphics[width=1\textwidth]{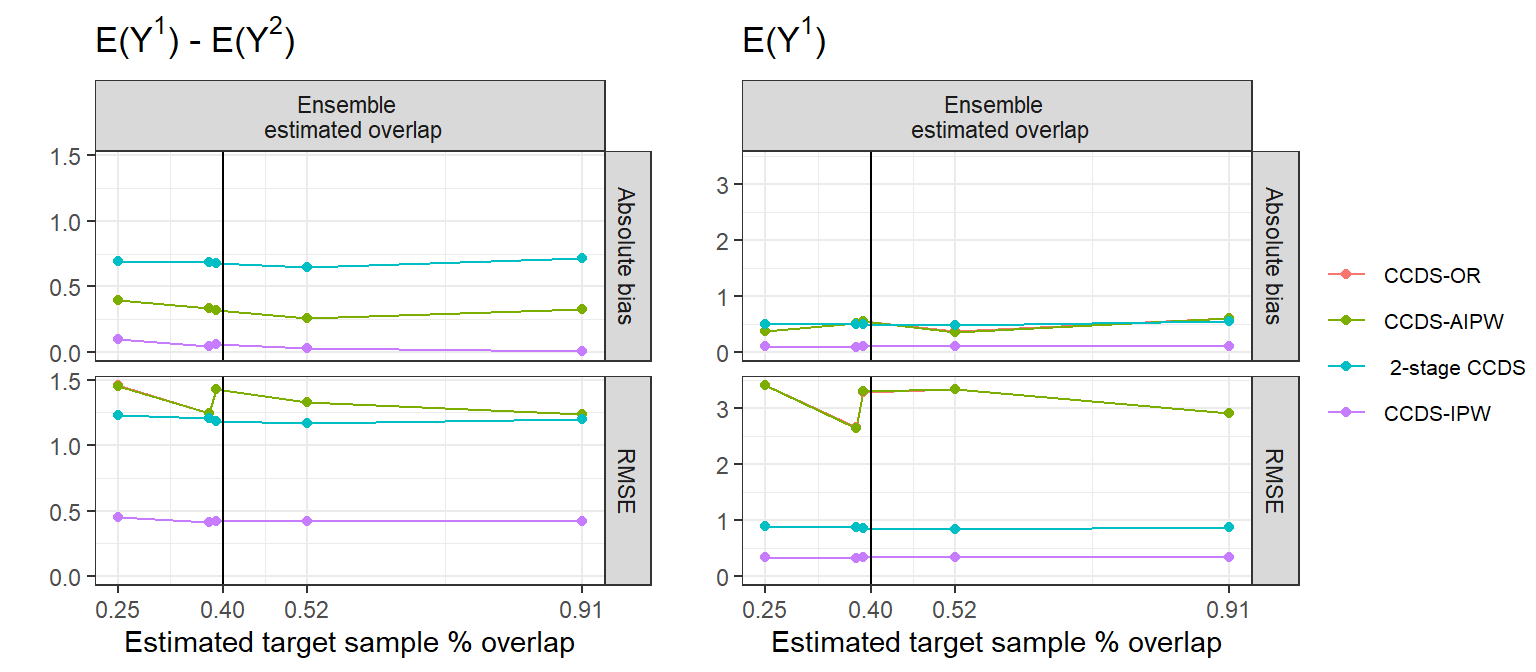}	
			\caption{Impact of Different Overlap Region Specifications on Bias and RMSE. The true percent overlap is distinguished by a vertical black line.}
			\label{figure:overlap_spec} 	
		\end{center}
	\end{figure}

	We examined a range of overlap region specifications (Table \ref{table:results_overlap_region}), with $\alpha$ and $\beta$ overlap region hyperparameters set on the propensity and log propensity scales. The overlap region specifications used in the base case ($\alpha=1\% \times \text{range}(\text{logit}(\pi_S))$ and $\beta=1\% \times \text{min}(n_\text{obs},n_\text{rand})$) were the closest to approximating the true overlap region, particularly for randomized data. Across the range of scenarios examined, which spanned underestimating to grossly overestimating the overlap region, bias and RMSE were minimally impacted (Figure \ref{figure:overlap_spec}).
	
	\begin{table}[]
		\centering
		\begin{tabular}{lllllllll}
			
			& \multicolumn{3}{c}{Mean Overlap}  \\ 
			& Obs             & RCT            & Total   \\ \hline
			Truth                                                                                 & 38\%             & 50\%            & 40\%         \\ \hline
			{$\begin{aligned}
					\alpha &= 1\% \times \text{range}(\pi_S) = 0.01 \\
					\beta &= 1\% \times \text{min}(n_\text{obs},n_\text{rand})) = 20 \end{aligned}$}       & 24\%             & 29\%            & 25\%       \\ \hline
			{$\begin{aligned}
					\alpha &= 1\% \times \text{range}(\text{logit}(\pi_S)) = 0.3 \\
					\beta &= 1\% \times \text{min}(n_\text{obs},n_\text{rand})) = 20 \end{aligned}$}       & 35\%             & 48\%            & 38\%       \\ \hline
			{$\begin{aligned}
					\alpha&=2\% \times \text{range}(\pi_S) = 0.02 \\
					\beta&=1\% \times \text{min}(n_\text{obs},n_\text{rand}) = 20 \end{aligned}$}
			& 38\%             & 42\%            & 39\%       \\ \hline
			{$\begin{aligned}
					\alpha &= 2\% \times \text{range}(\text{logit}(\pi_S)) = 0.6 \\
					\beta &= 1\% \times \text{min}(n_\text{obs},n_\text{rand}) = 20 \end{aligned}$}  & 50\%             & 61\%            & 52\%       \\ \hline
			{$\begin{aligned}
					\alpha &= 10\% \times \text{range}(\text{logit}(\pi_S)) = 3 \\
					\beta &= 4\% \times \text{min}(n_\text{obs},n_\text{rand}) = 78 \end{aligned}$} & 91\%             & 89\%            & 91\%          \\ \hline
			
		\end{tabular}
		\caption{Overlap Region Specifications.}
		\label{table:results_overlap_region}
	\end{table}

	\begin{figure}
		\spacingset{1}
		\begin{center}
			\includegraphics[width=1\textwidth]{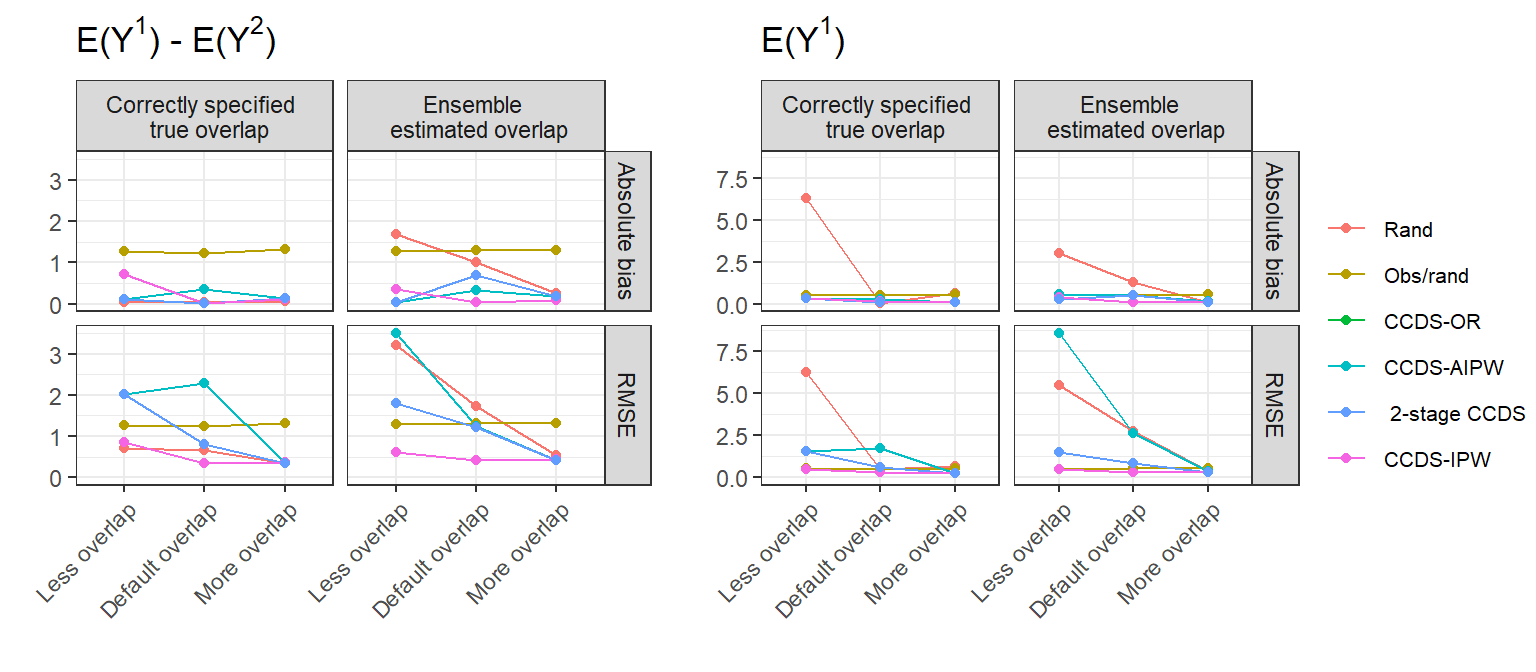}	
			\caption{Impact of Degree of Overlap (Positivity of Selection Violation) on Bias and RMSE. The settings examined include 13\% (less overlap), 38\% (default overlap), and 88\% (greater overlap) of observational data and 50\% of randomized data in the overlap region.}
			\label{figure:results_overlap}
		\end{center}
	\end{figure}
	
	\subsection{Different Degrees of Overlap (Positivity of Study Selection Violation)}
	We changed the size of $\mathcal{R}_\text{overlap}$ from the default of $QNorm(0.5)=0 \le X_1 \le 1.28 = QNorm(0.9)$ to $QNorm(0.7)=0.52 \le X_1 \le 1.28 = QNorm(0.9)$ for less overlap and $QNorm(0.1)=-1.28 \le X_1 \le 1.28 = QNorm(0.9)$ for greater overlap. These changes resulted in different proportions of observational data falling in the overlap region (13\%, 38\%, and 88\%, respectively), but always retained 50\% of randomized data in the overlap region.
	
	With greater overlap, all estimators besides obs/rand were able to shrink the bias close to zero (Figure \ref{figure:results_overlap}). Novel estimators had a larger region in which to estimate bias and the rand estimator was able to extrapolate better because more of the target population was in its region of support. With less overlap, all novel estimators' bias remained minimal, although variance increased (most starkly for the CCDS-OR and CCDS-AIPW estimators with ensembles); rand model bias increased sharply for estimating PTSMs. The data-generating mechanism allowed PATEs to be extrapolated from the randomized data, thus a corresponding bias increase was not observed for PATEs. The CCDS-IPW's RMSE remained the least impacted among all novel estimators.

	%
	%
	
	\subsection{Different Ratios of $n_\text{RCT}:n_\text{obs}$}
\begin{figure}
		\spacingset{1}
		\begin{center}
			\includegraphics[width=1\textwidth]{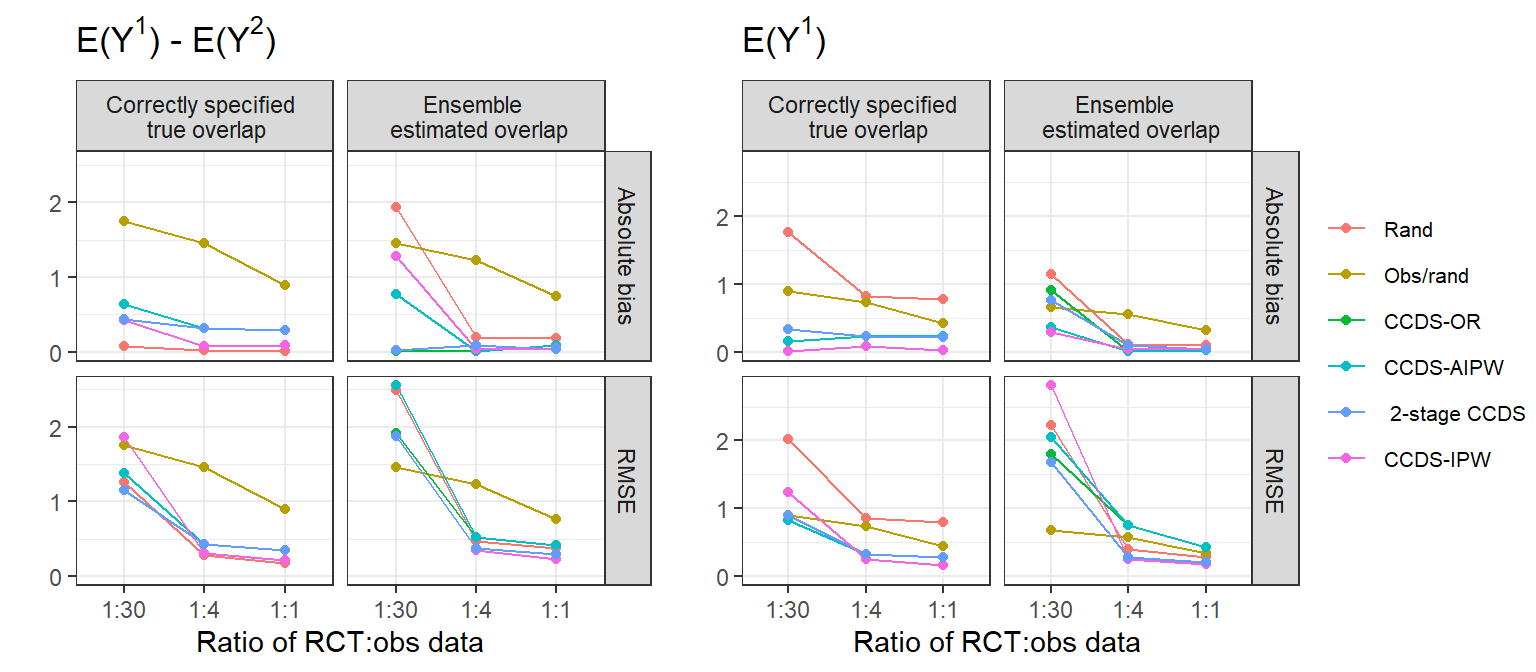}	
			\caption{Impact of Different Ratios of $n_\text{RCT}:n_\text{obs}$ on Bias and RMSE.}
			\label{figure:results_ratio_obs_rand}
		\end{center}
	\end{figure}

	The base-case ratio of $n_\text{RCT}:n_\text{obs}$ was 1:4. We also examined 1:1 and 1:30 ratios. To maintain the same overlap region across all settings, we changed the overlap region bounds to $QNorm(0.18)=-0.92 \le X_1 \le 2.05 = QNorm(0.98)$. Bias and RMSE largely decreased across all estimators as the ratio of randomized to observational  increased (Figure \ref{figure:results_ratio_obs_rand}). As the randomized observations comprised a larger portion of the target sample, the bias from using the rand and obs/rand estimators also  decreased. With ensembles, the rate of bias and RMSE decrease for rand and novel estimators exceeded that of the obs/rand estimator, highlighting the large impact of having more randomized data when overlap is small. With correct specification, the rate of RMSE decrease exceeded that of the obs/rand estimators. Relative performance  of the estimators largely remained the same.

	\subsection{Varying Sample Sizes} 
\begin{figure}
		\spacingset{1}
		\begin{center}
			\includegraphics[width=1\textwidth]{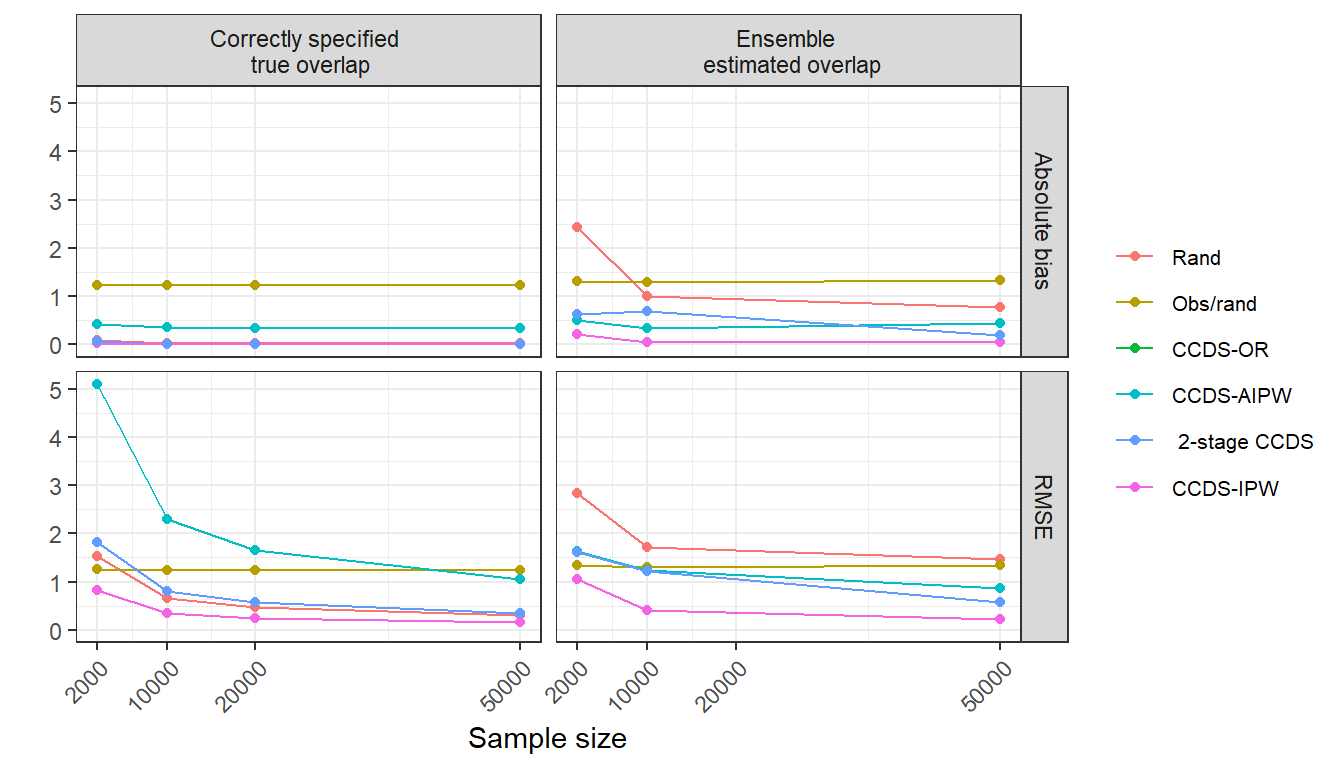}	
			\caption{Impact of $n$ on Bias and RMSE for PATE.}
			\label{figure:results_n}
		\end{center}
	\end{figure}

	We examined sample sizes $n=2000$, $10000$, and $50000$. 
	With correctly specified regressions, all other estimators had bias lower than that of the obs/rand estimator, although the CCDS-OR and CCDS-AIPW estimators retained some bias due to  fitting complex regressions in the small overlap region (Figure \ref{figure:results_n}). With smaller sample sizes ($n=2000$; $n_\text{RCT}=500$), the RMSE of all novel estimators and the rand estimator exceeded that of the obs/rand estimator. With ensembles, all novel estimators' bias was below that of the obs/rand estimator. RMSE, however, only dropped below that of the obs/rand estimator with $n=10000$ for the PATE (except for the CCDS-IPW estimator where RMSE was lower even with $n=2000$). 
	
	\subsection{Varying Strengths of Unmeasured Confounding} 
\begin{figure}
		\spacingset{1}
		\begin{center}
			\includegraphics[width=1\textwidth]{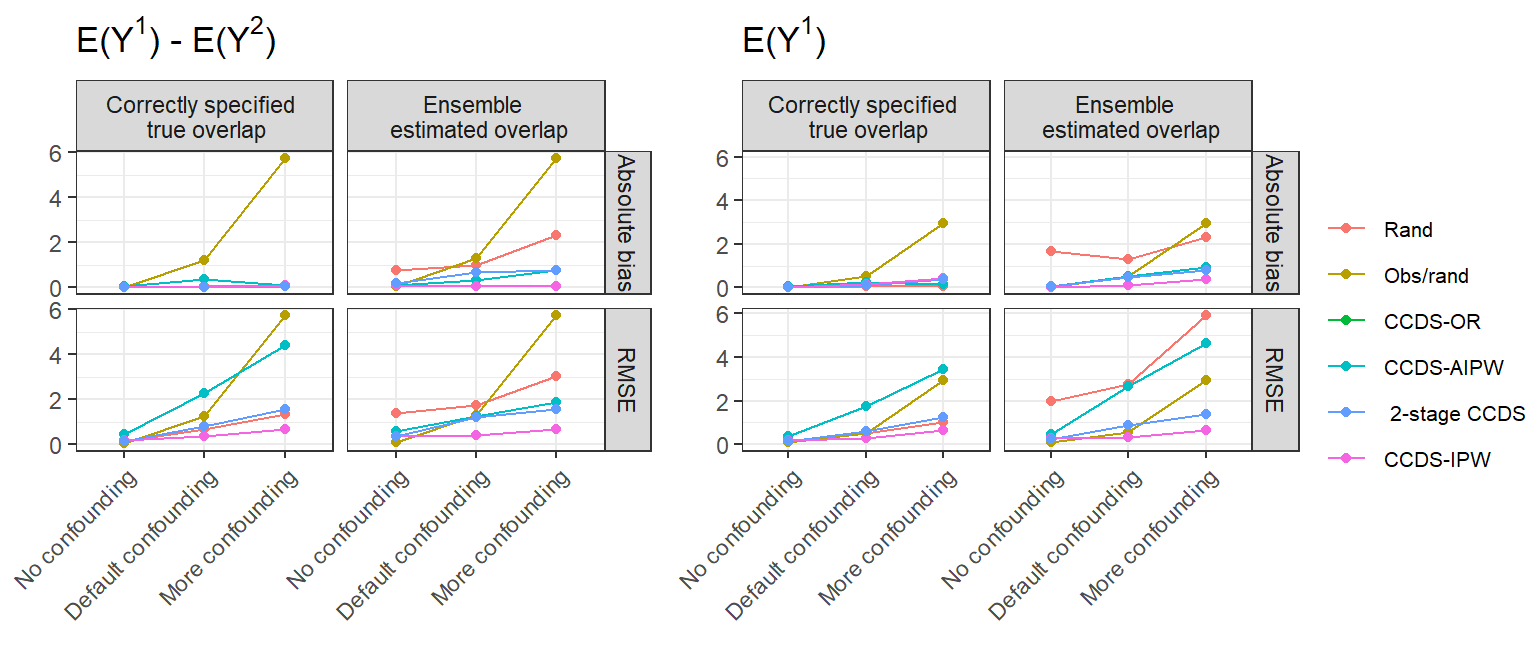}	
			\caption{Impact of Unmeasured Confounding on Bias and RMSE.}
			\label{figure:results_U}	
		\end{center}
	\end{figure}

	We examined four settings: no unmeasured confounding (where $U$ was included as a measured covariate) and three levels of unmeasured confounding. The coefficient for  $U$ was assigned three different values in  $P(Y|S,X,U)$:   (0.1, 0.625, 1.5) and in  $P(A|S,X,U)$: (5, 10,  20). These settings represented  low, default, and high confounding, respectively. Results were similar for no and low unmeasured confounding. As unmeasured confounding bias increased, with correctly specified regressions, there was no corresponding increase in bias across novel estimators. However,  variance increased, reflecting more uncertainty in settings with greater unmeasured confounding (Figure \ref{figure:results_U}). With ensembles, there was a small increase in bias with greater confounding. This increase was smaller for CCDS estimators than for the rand estimator.
	
	\subsection{Constant Conditional Bias Assumption Violation}
	To violate the constant conditional bias assumption, the amount of unmeasured confounding bias was varied in a way that is not predictable from the trends observed in the overlap region (the overlap region lower bound is at $X_1=0$): $E(Y^1) = E_\text{base}(Y^1) - 45U\times (X_1 + 0.5)\times I(X_1 < -0.5)$ and $E(Y^2) = E_\text{base}(Y^2) - 30U\times (X_1 + 0.5)\times I(X_1 < -0.5)$, where $E_\text{base}(Y^a)$ corresponds to the base-case potential outcome. 
		\begin{figure}
		\spacingset{1}
		\begin{center}
			\includegraphics[width=1\textwidth]{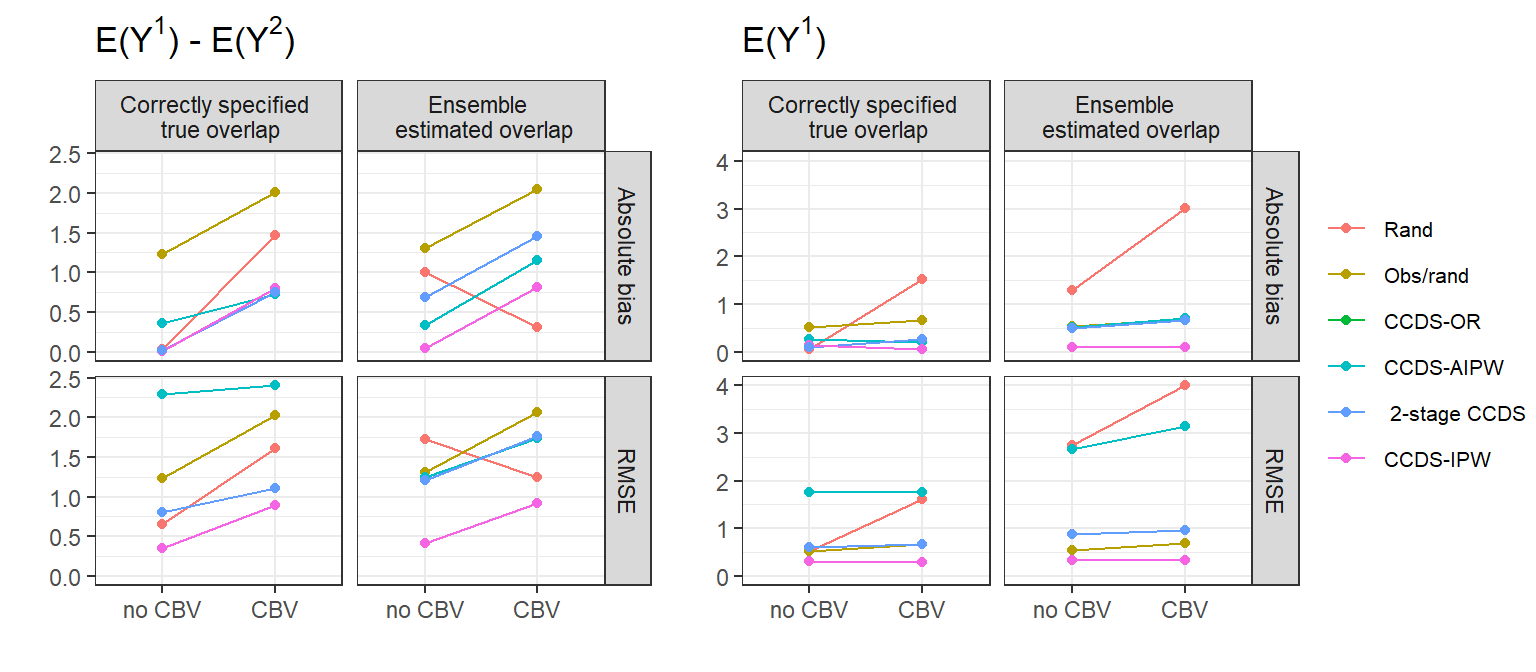}	
			\caption{Impact of Constant Conditional Bias Assumption Violation on Bias and RMSE. CBV is constant bias violation.}
			\label{figure:results_cbv}
		\end{center}
	\end{figure}
	When the bias relationship observed in the overlap region differed from  outside the overlap region, bias and RMSE increased for each of the estimators relative to the amount of extra unmeasured confounding bias  in the observational data that cannot be estimated from the overlap region (Figure \ref{figure:results_cbv}). The rand estimator was  not able to extrapolate well outside the overlap region even with a correctly specified regression. Novel estimators' bias remained below that of the obs/rand estimator as they removed the portion of unmeasured confounding bias that was estimable from the overlap region. The bias increase observed with constant conditional bias assumption violation thus accommodates the extra unmeasured confounding bias which cannot be removed, reflecting that these novel methods can only remove bias estimable from the overlap region; they rely on the constant conditional bias assumption. With ensembles, the rand estimator's PATE bias and RMSE decreased with the constant bias violation, likely reflecting this specific  data-generating mechanism as this result was not observed for PTSMs. When estimating PTSMs, the rand estimator was the most impacted by the assumption violation. 
	

	\subsection{Exchangeability of Study Selection Violation}
\begin{figure}
		\spacingset{1}
		\begin{center}
			\begin{subfigure}[t]{1\textwidth}
				\centering \includegraphics[width=1\textwidth]{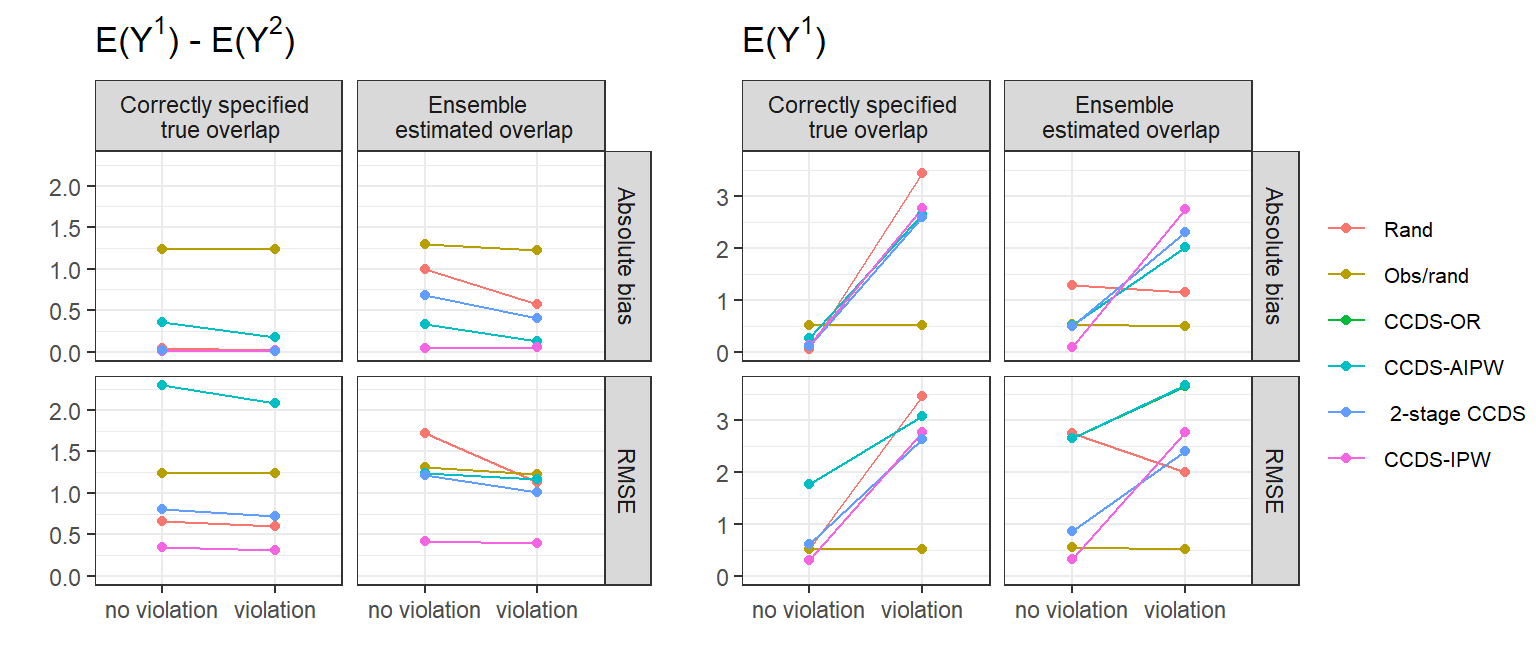}	 
			\end{subfigure}	
			\textbf{(a)} $S$ function of $U$
			
			\begin{subfigure}[t]{1\textwidth}
				\centering \includegraphics[width=1\textwidth]{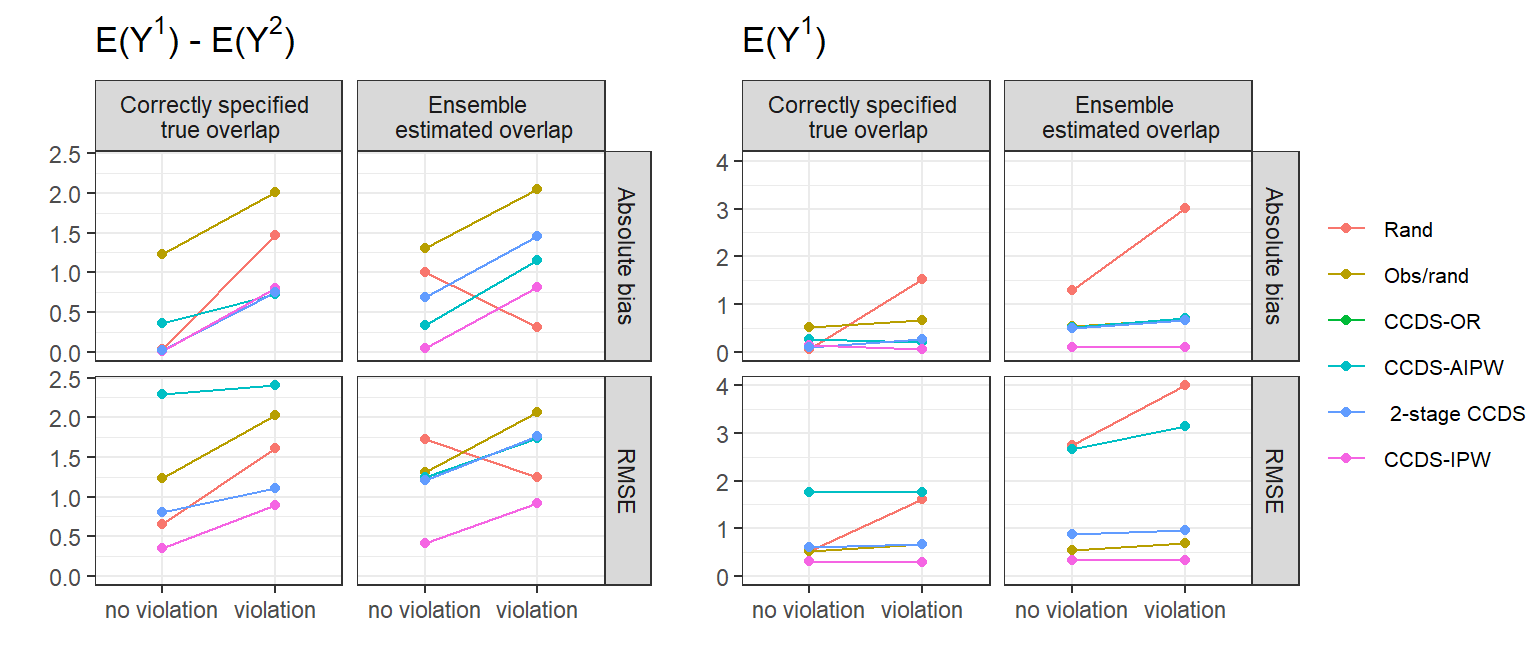}	 
			\end{subfigure}
			\textbf{(b)} $U$ function of $X_1$
			\caption{Impact of Exchangeability of Study Selection Assumption Violation on Bias and RMSE.}
			\label{figure:results_exchangeability_SU}
		\end{center}
	\end{figure}

	We examined violations of the exchangeability of study selection assumption  through two approaches. In the first, $P(S|\bm{X},U)$ was changed to be a function of $U$ in the overlap region, $P(S=1|U)=0.125+0.25U$, and remained deterministically 0 or 1 outside the overlap region. $P(S=1)$ remained at $0.20$. When study selection was a function of unmeasured $U$, bias for PTSMs of all other estimators exceeded that of the obs/rand estimator (Figure \ref{figure:results_exchangeability_SU}). This was not observed for the PATEs due to $U$ not being an unmeasured effect modifier.
	
	
	In the second violation assessment, $U$ was a function of $X_1$, which determines overlap region membership: $U \sim Binom(p_U)$, where $p_U = \text{expit}(30X_1)$. Hence, $U$ was an unmeasured effect modifier with different distributions in the randomized vs. observational data. Thus, randomized estimates would not represent the truth in the overlap region. Hence, the CCDS estimators' bias increased with the assumption violation; when rand estimates are biased for the target population quantities in the overlap region, CCDS estimators are not be able to properly debias (Figure \ref{figure:results_exchangeability_SU}). Likewise, the rand estimates' bias also increased in all but the ensemble estimating the PATE, where  bias  decreased, likely due to bias cancellation between treatment groups.

	\subsection{Probabilistic $\pi_S$}\label{appendix:simulation_s_probabilistic}
\begin{figure}
		\spacingset{1}
		\begin{center}
			\includegraphics[width=0.5\textwidth]{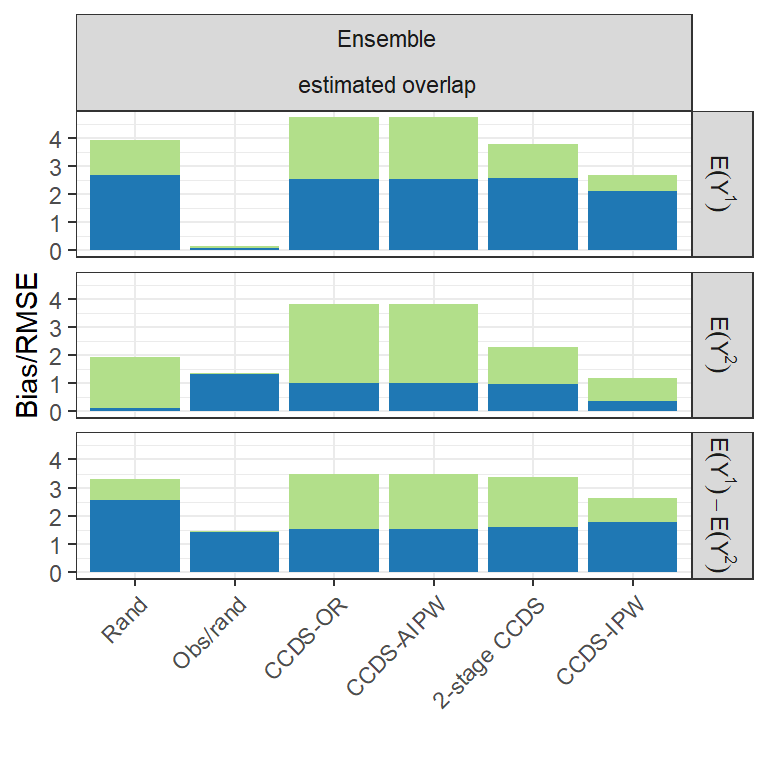}	 
			
			\caption{Impact of Probabilistic $\pi_S$ and Smaller $P(A=1)$ on Bias and RMSE.}
			\label{figure:results_S_probabilistic}
		\end{center}
	\end{figure}

	For a probability of selection that was (more realistically) probabilistically rather than deterministically based on $X_1$, we generated $S \sim Binom(P_S)$, where $\text{logit}(P_S) = \beta_S X$. Namely, $\beta_S = (-\beta_{S0},2,1,0.5,0.1,0.1,0.1)$, where $X = (1, X_1, X_2, X_3, X_4, X_1^3, \text{sine}(X_1^3*X_2))$ and $\beta_{S0}$ is set to ensure $P(S)=0.2$. We also set $P(A|S=1) = 0.1$ and changed the intercept of the logistic regression determining $A$ for $S=0$ to $-3.5$ such that $P(A|S=0) \approx 0.1$. As a result, estimation of $E(Y^1)$ using CCDS-IPW becomes more difficult, and CCDS-IPW's bias becomes on par with that of other CCDS estimators (Figure \ref{figure:results_S_probabilistic}).
	
	\subsection{Alternative Data Generating Mechanisms}
	We examined alternative data-generating mechanisms for $Y$, $S$, and $A$. This included using simple main terms linear forms (excluding $X_1^3$ terms), including more complex effect heterogeneity such that the PATE was not extrapolatable from the randomized data ($\mu_Y=-1.5-3A+4X_1+4X_2+3X_3+2X_4+2(X_1+1)^3+4AX_1+2A(X_1+1)^3+10U$), using knot terms ($\mu_Y = \mu_\text{Y, base} - 15I(X_1 < -1)(X_1 + 1) - 15I(X_1 < -1)(X_1 + 1)A$, $\mu_Y = \mu_\text{Y, base} - 45I(X_1 < -0.5)(X_1 + 0.5) + 15I(X_1 < -0.5)(X_1 + 0.5)A$, $\mu_Y = \mu_\text{Y, base} - 2I(X_1 < 0.5)(X_1 - 0.5) - 2I(X_1 < 0.5)(X_1 - 0.5)A$), and including an interaction between the unmeasured confounder and a measured covariate ($\mu_Y = \mu_\text{Y, base} + 2UX_1 + UX_1A$), where $\mu_\text{Y, base}$ is the outcome mean in the base case. Conclusions remained similar across all examined data-generating mechanisms. 
	
	
	\section{Supplemental Medicaid Results}\label{appendix:Medicaid}
	
\begin{figure}
		\spacingset{1}
		\begin{center}
			\includegraphics[width=0.55\textwidth]{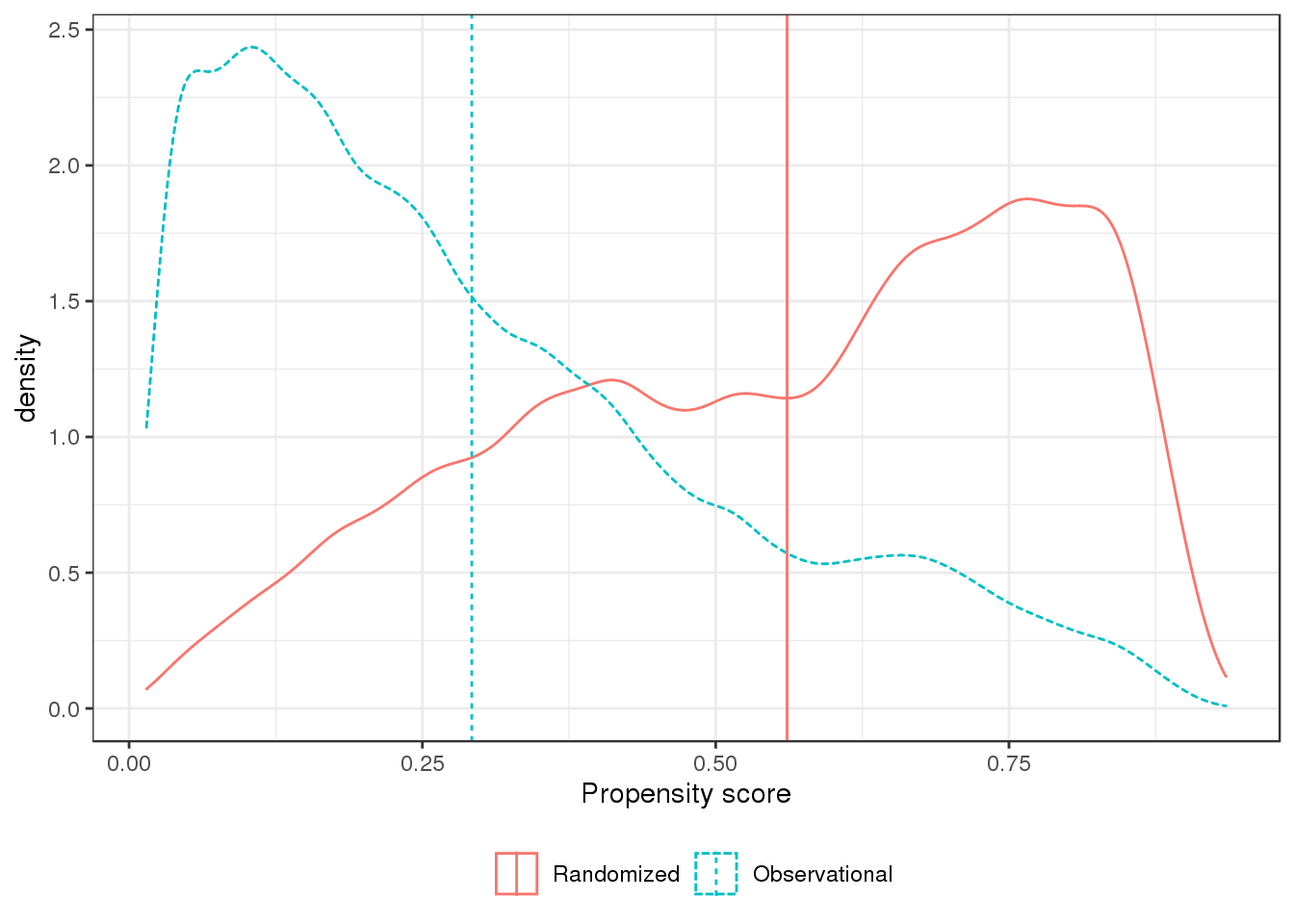}	
			\caption{Propensity for Selection into the Randomized Group. The plot displays the density and mean (vertical line), estimated using a linear regression.}
			\label{figure:results_Medicaid_piS}
		\end{center}
	\end{figure}

\begin{figure}
		\spacingset{1}
		\begin{center}
			\includegraphics[width=0.7\textwidth]{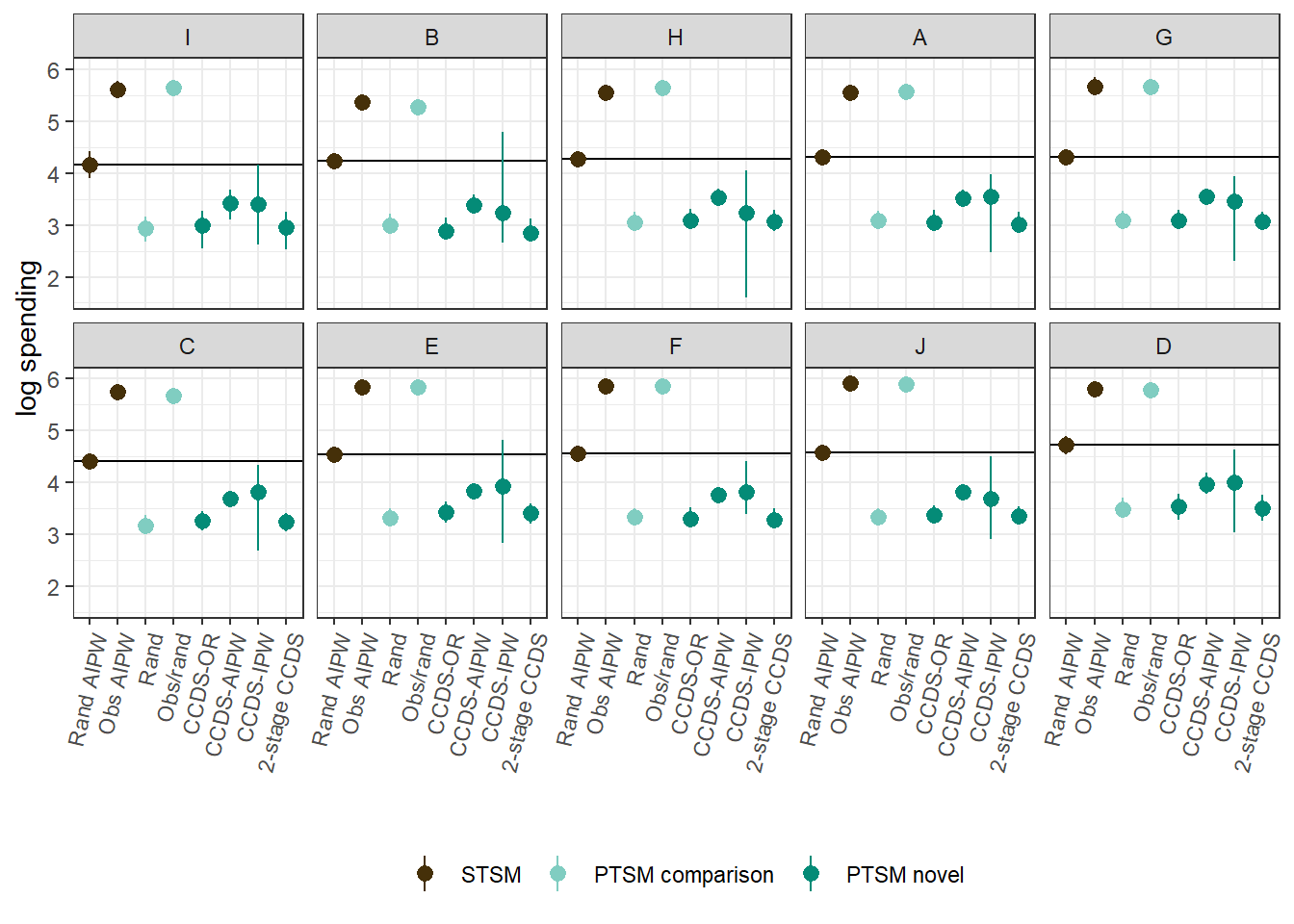}	
			\caption{STSMs and PTSMs Across Health Plans for All Estimators, with 95\% Confidence Intervals Multiplicity-Adjusted With the Bonferroni Correction.}
			\label{figure:results_Medicaid_all}
		\end{center}
	\end{figure}

	Summary characteristics of the randomized and observational groups in the Medicaid data are presented in Table \ref{table:MedicaidTable1}. Plots of propensity for selection into the randomized group for both the randomized and observational data are displayed in Figure \ref{figure:results_Medicaid_piS}. STSMs and PTSMs for all health plans and estimators can be found in Figure \ref{figure:results_Medicaid_all}.
	
	\begin{table}
		\scriptsize		
		\begin{tabular}{llccc}
			\hline
			\multicolumn{2}{l}{\textbf{Characteristic}}                                              & \textbf{Randomized}              & \textbf{Observational}              & \textbf{p-value}                   \\ \hline
			\multicolumn{2}{l}{Sample size}                                                          & 65591                            & 98232                               &                              \\
			\multicolumn{2}{l}{6 month spending   (mean (SD))}                                       & 3052 (10089)                     & 2796 (6756)                         & \textless{}0.001             \\
			\multicolumn{2}{l}{Plan (n (\%))}                                                        &                                  &                                     & \textless{}0.001             \\
			& A                                                          & 8510 ( 13.0)                     & 9879 ( 10.1)                        &                              \\
			& B                                                          & 7814 ( 11.9)                     & 6390 ( 6.5)                         &                              \\
			& C                                                          & 6195 ( 9.4)                      & 6200 ( 6.3)                         &                              \\
			& D                                                          & 2626 ( 4.0)                      & 18149 ( 18.5)                       &                              \\
			& E                                                          & 6770 ( 10.3)                     & 11302 ( 11.5)                       &                              \\
			& F                                                          & 8055 ( 12.3)                     & 17673 ( 18.0)                       &                              \\
			& G                                                          & 8439 ( 12.9)                     & 5689 ( 5.8)                         &                              \\
			& H                                                          & 7062 ( 10.8)                     & 6833 ( 7.0)                         &                              \\
			& I                                                          & 1420 ( 2.2)                      & 3402 ( 3.5)                         &                              \\
			& J                                                          & 8700 ( 13.3)                     & 12715 ( 12.9)                       &                              \\
			\multicolumn{2}{l}{Age (mean (SD))}                                                      & 35.55 (12.65)                    & 34.26 (12.75)                       & \textless{}0.001             \\
			\multicolumn{2}{l}{Female (n (\%))}                                                      & 26370 ( 40.2)                    & 58076 ( 59.1)                       & \textless{}0.001             \\
			\multicolumn{2}{l}{County (n (\%))}                                                      &                                  &                                     & \textless{}0.001             \\
			& Bronx                                                      & 16423 ( 25.0)                    & 21942 ( 22.3)                       &                              \\
			& Brooklyn                                                   & 21044 ( 32.1)                    & 32307 ( 32.9)                       &                              \\
			& Manhattan                                                  & 13281 ( 20.2)                    & 13002 ( 13.2)                       &                              \\
			& Queens                                                     & 12679 ( 19.3)                    & 27544 ( 28.0)                       &                              \\
			& Staten Island                                              & 2164 ( 3.3)                      & 3437 ( 3.5)                         &                              \\
			\multicolumn{2}{l}{Aid group (n (\%))}                                                   &                                  &                                     & \textless{}0.001             \\
			& MA SN adult                                                & 31430 ( 47.9)                    & 56210 ( 57.2)                       &                              \\
			& MA SN child                                                & 102 ( 0.2)                       & 339 ( 0.3)                          &                              \\
			& MA SSI blind                                               & 714 ( 1.1)                       & 431 ( 0.4)                          &                              \\
			& MA TANF adult                                              & 10867 ( 16.6)                    & 27553 ( 28.0)                       &                              \\
			& MA TANF child                                              & 931 ( 1.4)                       & 2246 ( 2.3)                         &                              \\
			& SN adult                                                   & 15573 ( 23.7)                    & 6267 ( 6.4)                         &                              \\
			& SN child                                                   & 99 ( 0.2)                        & 125 ( 0.1)                          &                              \\
			& SSI blind                                                  & 5114 ( 7.8)                      & 1358 ( 1.4)                         &                              \\
			& TANF adult                                                 & 648 ( 1.0)                       & 3520 ( 3.6)                         &                              \\
			& TANF child                                                 & 65 ( 0.1)                        & 146 ( 0.1)                          &                              \\
			& Other                                                      & 48 ( 0.1)                        & 37 ( 0.0)                           &                              \\
			\multicolumn{2}{l}{Eligible for SSI   (n (\%))}                                          & 5840 ( 8.9)                      & 1797 ( 1.8)                         & \textless{}0.001             \\
			\multicolumn{2}{l}{Baseline spending   decile (mean (SD))}                               & 6.24 (3.31)                      & 3.88 (3.40)                         & \textless{}0.001             \\
			\multicolumn{2}{l}{Missing baseline   spending (n (\%))}                                 & 839 ( 1.3)                       & 715 ( 0.7)                          & \textless{}0.001             \\
			\multicolumn{2}{l}{Percent   neighborhood poverty (mean (SD))}                           & 0.24 (0.08)                      & 0.23 (0.08)                         & \textless{}0.001             \\ \hline
			\multicolumn{5}{p{\textwidth}}{\textit{NOTE: The   p-values correspond to a t-test for continuous variables and a chi-squared   test for categorical variables, with a continuity correction.}}                    \\
			\multicolumn{5}{p{\textwidth}}{\textit{Abbreviations:   MA = Medicare Advantage; SD = standard deviation; SN = safety net; SSI =   social security income; TANF = Temporary Assistance for Needy Families.}}
		\end{tabular}
		\caption{Characteristics of Randomized and Observational Medicaid Groups.}
		\label{table:MedicaidTable1}
		
	\end{table}

	\subsection{Accounting for Country-Month-Year} \label{appendix:CMY}

	Due to computational considerations, the primary analysis of the Medicaid study did not account for correlation by country-month-year, the unit of randomization. In sensitivity analyses, we accounted for this correlation using either (1) fixed effects, similar in spirit to the analysis by \cite{geruso2020}, or (2) random effects. Analyses were run paralleling the linear rand estimator, i.e., fitting a linear outcome regression on the randomized data. We recovered similar estimates for the randomized population and slightly higher point estimates for the target population compared to the primary analysis (Table \ref{table:MedicaidCMY}). These correspond to a 67\% (fixed effects estimate) and 63\% (random effects estimate) difference between 6-month spending for the randomized population had it been in the highest- vs. lowest-spending plans and a 67\% and a 65\% difference, respectively, for the target population. In comparison, when not accounting for correlation by country-month-year, linear regression versions of the main analysis found a 70\% difference for the randomized population and 75\% difference for the target population.

			\begin{table}[]
		\spacingset{1}
		\begin{tabular}{lllll}
			\hline
			Sample     & Plan & \begin{tabular}[c]{@{}l@{}}CMY\\ fixed effects\end{tabular} & \begin{tabular}[c]{@{}l@{}}CMY\\ random effects\end{tabular} & \begin{tabular}[c]{@{}l@{}}No CMY \\ (primary analysis, \\ linear version)\end{tabular} \\ \hline
			Randomized & I    & 4.25                                                        & 4.22                                                         & 4.19                                                                                 \\
			& D    & 4.76                                                        & 4.71                                                         & 4.72                                                                                 \\
			Target     & I    & 3.42                                                        & 3.38                                                         & 2.92                                                                                 \\
			& D    & 3.93                                                        & 3.88                                                         & 3.48                                                                                 \\ \hline                                                                                                                                                             
		\end{tabular}
		\caption{STSMs and PTSMs Across Analyses with Differing Correlation Procedure for Country-Month-Year (CMY). All analyses fit a linear outcome regression using the randomized data.}
		\label{table:MedicaidCMY}
	\end{table}

	\pagebreak
	
	\bibliographystyle{Chicago}
	\bibliography{CCDS_for_generalizability.bib}